\RequirePackage{ifpdf}
\ifpdf % We are running pdfTeX in pdf mode
\documentclass[pdftex]{sigma}
\else
\documentclass{sigma}
\fi

\numberwithin{equation}{section}

\begin{document}

\allowdisplaybreaks

\renewcommand{\thefootnote}{$\star$}

\renewcommand{\PaperNumber}{051}

\FirstPageHeading

\ShortArticleName{Generic Superintegrable System on the 3-Sphere}

\ArticleName{Two-Variable Wilson Polynomials and the Generic \\ Superintegrable System on the 3-Sphere\footnote{This paper is a
contribution to the Special Issue ``Relationship of Orthogonal Polynomials and Special Functions with Quantum Groups and Integrable Systems''. The
full collection is available at
\href{http://www.emis.de/journals/SIGMA/OPSF.html}{http://www.emis.de/journals/SIGMA/OPSF.html}}}

\Author{Ernie G.~KALNINS~$^\dag$,  Willard  MILLER  Jr.~$^\ddag$  and Sarah POST~$^\S$}

\AuthorNameForHeading{E.G.~Kalnins,   W.~Miller  Jr.\ and S.~Post}

\Address{$^\dag$~Department of Mathematics,  University of Waikato, Hamilton, New Zealand}
\EmailD{\href{mailto:math0236@math.waikato.ac.nz}{math0236@math.waikato.ac.nz}}
\URLaddressD{\url{http://www.math.waikato.ac.nz}}

\Address{$^\ddag$~School of Mathematics, University of Minnesota,
 Minneapolis, Minnesota, 55455, USA}
\EmailD{\href{mailto:miller@ima.umn.edu}{miller@ima.umn.edu}}
\URLaddressD{\url{http://www.ima.umn.edu/~miller/}}

\Address{$^\S$~Centre de Recherches Math\'ematiques,
Universit\'e de Montr\'eal, \\
\hphantom{$^\S$}~C.P. 6128 succ. Centre-Ville, Montr\'eal (QC) H3C 3J7, Canada}
\EmailD{\href{mailto:sarahisabellepost@gmail.com}{sarahisabellepost@gmail.com}}
\URLaddressD{\url{http://www.crm.umontreal.ca/~post/}}

\ArticleDates{Received January 31, 2011, in f\/inal form May 23, 2011;  Published online May 30, 2011}

\Abstract{We show that the symmetry operators for the quantum superintegrable system on the 3-sphere with
generic 4-parameter  potential form a closed quadratic algebra with 6 linearly independent generators that closes
at order 6 (as dif\/ferential operators). Further there is an algebraic relation at order 8 expressing the fact that
there are only 5 algebraically independent generators. We  work out the details of modeling physically relevant
irreducible representations of the quadratic algebra in terms of divided dif\/ference operators in two variables.
We determine several ON bases for this model including  spherical and  cylindrical bases. These bases are expressed
in terms of two variable Wilson and Racah polynomials with arbitrary parameters, as def\/ined by Tratnik.
The generators for the quadratic algebra are expressed in terms of
recurrence operators for the one-variable Wilson polynomials. The quadratic algebra structure breaks the degeneracy
of the space of these polynomials. In an earlier paper the authors found a similar characterization of one variable
Wilson and Racah polynomials in terms of irreducible representations of the quadratic algebra for the quantum
superintegrable system on the 2-sphere with generic 3-parameter  potential. This indicates a general relationship
between 2nd order superintegrable systems and discrete orthogonal polynomials.}


\Classification{81R12; 33C45}

\section{Introduction}\label{int1}

{\sloppy \looseness=1
We def\/ine  an $n$-dimensional classical superintegrable system to be  an integrable Hamiltonian
system that not only possesses $n$ mutually Poisson -- commuting constants of the motion, but in addition,
the Hamiltonian Poisson-commutes with $2n-1$ functions on the phase space that are globally
def\/ined and polynomial in the momenta. Similarly, we def\/ine a quantum
superintegrable system to be a quantum Hamiltonian which is one of a set
of $n$ algebraically independent mutually commuting dif\/ferential operators, and that commutes with a set of $2n-1$
independent dif\/fe\-ren\-tial operators of f\/inite order. We restrict to classical systems of the
form ${\cal H}=\sum\limits_{i,j=1}^ng^{ij}p_ip_j+V$ and  quantum
systems $H=\Delta_n+{\tilde V}$. These systems, inclu\-ding the classical Kepler \cite{CORDANI} and anisotropic oscillator systems
and the quantum anisotropic oscillator and hydrogen atom have great historical importance, due to their
remarkable pro\-per\-ties \cite{SCQS,IMA, Higgs,CZ1, CZ2}. One modern practical application among many  is the Hohmann transfer, a~fundamental tool
for the positioning of earth satellites and for
celestial navigation in ge\-ne\-ral, which is based on the superintegrability of the Kepler system \cite{CURTIS}. The order of a classical
superintegrable system is the maximum order
of the generating constants of the motion (with the Hamiltonian excluded) as a polynomial in the momenta,
and the order of a quantum superintegrable system is the
 maximum order of the quantum symmetries as dif\/ferential operators.

 }

 Systems of 2nd order have been well studied and
there is now a structure and classif\/ication theory \cite{KKM20061,KKMP, DASK2005, KKM2007,KMP2007,KMP2008},
especially for the cases $n=2,3$.
For 3rd and higher order superintegrable systems there have been recent dramatic advances but no structure and
classif\/ication theory as yet
\cite{CDR2008, RTW, Evans2008a,TTW,TTW2,CQ10,TW2010, PW2010, KMP10,KKM10,KKM10a,KMPog10, BH, Marquette1, Marquette2, RRS}.

{\sloppy \looseness=1
The potential $V$ corresponding to a 2nd order superintegrable system, classical or quantum, on an $n$-dimensional
conformally f\/lat manifold depends linearly on several parameters in general and can be shown to generate a
vector space of dimension $\le n+2$. (One dimension corresponds to the trivial addition of a constant to the
potential and usually isn't included in a parameter count.) If the maximum is achieved, the potential is called nondegenerate. There is an invertible mapping between
superintegrable systems on dif\/ferent manifolds, called the St\"ackel transform, which preserves the structure of the algebra ge\-ne\-rated
by the  symmetries. In the cases $n=2,3$ it is known that all nondegenerate 2nd order superintegrable systems are St\"ackel equivalent to a system on a
constant curvature space~\cite{KKM20042, KKM20052}. An important fact for 2D systems is that all systems can be obtained
from one  generic superintegrable system  on the complex
2-sphere by appropriately chosen limit processes, e.g.~\cite{KMR,KKWMPOG}. The use
of these processes in separation of variables methods for wave and
Helmholtz equations in $n$ dimensions was pioneered by
B\^ocher~\cite{Bocher}. For $n=3$ it appears that all nondegenerate 3D systems can be obtained
from one  generic superintegrable system  on the complex
3-sphere by similar limiting processes, but the proof is not yet complete~\cite{KKM2007, KKMP2011}.

}

For $n=2$ we def\/ine the  generic sphere system by embedding of the unit 2-sphere $x_1^2+x_2^2+x_3^2=1$
in three dimensional f\/lat space. Then the Hamiltonian operator is
\begin{gather*}%\label{qham2}
H=\sum_{1\le i<j\le 3}(x_i\partial_j-x_j\partial_i)^2+\sum_{k=1}^3\frac{a_k}{x_k^2}, \qquad
\partial_i\equiv\partial_{x_i}.
\end{gather*} The 3 operators that generate the symmetries are $L_1=L_{12}$, $L_2=L_{13}$, $L_3=L_{23}$ where
\begin{gather*}%\label{q2ordsym2}
L_{ij}\equiv L_{ji}=(x_i\partial_j-x_j\partial_i)^2+\frac{a_i x_j^2}{x_i^2}+\frac{a_j x_i^2}{x_j^2},
\end{gather*}
for $1\le i<j\le 4$.
Here
\[
H =\sum_{1\le i< j\le 3} L_{ij}+\sum_{k=1}^3a_k=H_0+V, \qquad V=\frac{a_1}{x_1^2}+\frac{a_2}{x_2^2}+\frac{a_3}{x_3^2}.
\]

\looseness=1
From the general structure theory for 2D 2nd order superintegrable systems with nondegenerate potential we know that the
3 def\/ining symmetries will generate a symmetry algebra (a~quadratic algebra)
by taking operator commutators, which closes at order~6,~\cite{KKM20041}. That is, all possible symmetries can be written as
symmetrized operator polynomials in the basis generators
and in the 3rd order commutator $R$, where $R$ occurs at most linearly. In particular, the dimension of the space of truly 2nd order
symmetries for the Hamiltonian operator  is~3, for the
3rd order symmetries it is 1, for the 4th order symmetries it is~6, and for the 6th order symmetries it is~10.
For the generic 2-sphere quantum system the structure equations
can be put in the symmetric form \cite{KMP2007}
\begin{gather}\label{structure1}\epsilon_{ijk} [L_i,R]=4\{L_i,L_k\}-4\{L_i,L_j\}- (8+16a_j)L_j + (8+16a_k)L_k+ 8(a_j-a_k),\\
 R^2=\frac83\{L_1,L_2,L_3\} -(16a_1+12)L_1^2 -(16a_2+12)L_2^2  -(16a_3+12)L_3^2\nonumber\\
\phantom{R^2=}{} +\frac{52}{3}(\{L_1,L_2\}+\{L_2,L_3\}+\{L_3,L_1\})+ \frac13(16+176a_1)L_1 +\frac13(16+176a_2)L_2 \nonumber\\
\phantom{R^2=}{}
+ \frac13(16+176a_3)L_3 +\frac{32}{3}(a_1+a_2+a_3) +48(a_1a_2+a_2a_3+a_3a_1)+64a_1a_2a_3. \label{structure2}
\end{gather}
Here $\epsilon_{ijk}$
is the pure skew-symmetric tensor,  $R=[L_1,L_2]$ and
$\{L_i,L_j\}=L_iL_j+L_jL_i$ with an analogous def\/inition of
$\{L_1,L_2,L_3\}$ as a symmetrized sum of 6 terms.
In practice we will substitute $L_3=H-L_1-L_2-a_1-a_2-a_3$ into these equations.

{\sloppy In \cite{KMP2007}
 we started from f\/irst principles and  worked out some families of f\/inite and inf\/inite dimensional irreducible
representations of the quadratic algebra  with structure rela\-tions~(\ref{structure1}),~(\ref{structure2}),
including those that corresponded to the bounded states of the associated quantum mechanical problem on the 2-sphere.
Then we found 1-variable models of these  representations in which the generators $L_i$ acted as divided dif\/ference operators
 in the variable~$t$
on a space of polynomials in~$t^2$. The eigenfunctions of one of the operators~$L_i$ turned out to be the
  Wilson  and Racah polynomials in their full generality. In essence, this described an isomorphism between the quadratic algebra of the generic quantum superintegrable
system on the 2-sphere and the quadratic algebra generated by the Wilson polynomials.

}

{\sloppy \looseness=1
The present paper is concerned with the extension of these results to the 3-sphere, where the situation is much more complicated.
From the general structure theory for 3D 2nd order superintegrable systems with nondegenerate potential we know that
although there are $2n-1=5$ algebraically independent 2nd order generators, there must exist a 6th 2nd order symmetry such that the
 6 symmetries are linearly independent and generate a
quadratic algebra that closes at order 6~\cite{KKM3}. (We call this the $5\Longrightarrow 6$ Theorem.)
  Thus, all possible symmetries can be written as symmetrized operator polynomials in the basis generators
and in the four 3rd order commutators~$R_i$, where the~$R_i$ occur at most linearly. In particular,   the dimension of the space of truly 2nd order symmetries is 3, for the
3rd order symmetries  is 4, for the 4th order symmetries it is 21, and for the 6th order sym\-metries it is~56.
 In 3D there are 5 algebraically independent, but 6
linearly independent, ge\-nerators. The algebra again closes at 6th
order, but in addition there is an identity at 8th
order that relates the 6 algebraically dependent generators. The
representation theory of such quadratic algebras is much more
complicated and we work out a very important instance of it here. In this case we will f\/ind an intimate relationship
between these representations and Tratnik's 2-variable Wilson and Racah polynomials in their full genera\-lity~\mbox{\cite{T1,T2,GI}}.

}

\looseness=1
 For $n$D  nondegenerate systems there are $2n-1$ functionally independent
but $n(n+1)/2$ linearly independent generators for the quadratic
algebra. We expect that the relationships developed here will extend to $n$-spheres although the results will be of increasing comple\-xity.

\section{The quantum superintegrable system on the 3-sphere}

We def\/ine the Hamiltonian operator via the embedding of the unit 3-sphere $x_1^2+x_2^2+x_3^2+x_4^2=1$
in four-dimensional f\/lat space
\begin{gather}\label{qham}
H=\sum_{1\le i<j\le 4}(x_i\partial_j-x_j\partial_i)^2+\sum_{k=1}^4\frac{a_k}{x_k^2}, \qquad
\partial_i\equiv\partial_{x_i}.
\end{gather} A basis for the second order constants of the motion is
\begin{gather*}%\label{q2ordsym}
L_{ij}\equiv L_{ji}=(x_i\partial_j-x_j\partial_i)^2+\frac{a_i x_j^2}{x_i^2}+\frac{a_j x_i^2}{x_j^2},
\end{gather*}
for $1\le i<j\le 4$.
Here
\[
H =\sum_{1\le i< j\le 4} L_{ij}+\sum_{k=1}^4a_k.
\]

In the following $i$, $j$, $k$, $\ell$ are pairwise distinct integers such that $1\le i,j,k,\ell\le 4$, and  $\epsilon_{ijk}$ is the completely skew-symmetric tensor
 such that $\epsilon_{ijk}=1$ if  $i<j<k$.
There are 4 linearly independent commutators of the second order
symmetries (no sum on repeated indices):
\begin{gather*}%\label{q3ordsym}
R_\ell=\epsilon_{ijk}[L_{ij},L_{jk}].
\end{gather*}
This implies, for example, that
\[
R_1=[L_{23},L_{34}]=-[L_{24},L_{34}]=-[L_{23},L_{24}].
\]
Also
\[
[L_{ij},L_{k\ell}]=0.
\]
Here we def\/ine the commutator of linear operators $F$, $G$ by
$[F,G]=FG-GF$. The structure equations can be worked out via a relatively straightforward but tedious process. We get the following results.

 The fourth order structure equations are
\begin{gather*}
%\label{q4ordsym1}
 [L_{ij},R_j]=4\epsilon_{i\ell k}(\{L_{ik},L_{j\ell}\}-\{L_{i\ell},L_{jk}\}+L_{i\ell}-L_{ik}+L_{jk}-L_{j\ell}),
\\
%\label{q4ordsym2}
[L_{ij},R_k]=4\epsilon_{ij\ell}(\{L_{ij},L_{i\ell}-L_{j\ell}\}+(2+4a_j)L_{i\ell}-(2+4a_i)L_{j\ell}+2a_i-2a_j).
\end{gather*}
Here $\{F,G\}=FG+GF$.

The f\/ifth order structure equations (obtainable directly from the
fourth order equations and the Jacobi identity) are
\begin{gather*}%\label{q5ordsym}
[R_\ell,R_k]=4\epsilon_{ik\ell}(R_i-\{L_{ij},R_i\}) +4\epsilon_{jk\ell}(R_j-\{L_{ij},R_j\}).\end{gather*}

The sixth order structure equations are
\begin{gather*}
R_\ell^2=\frac83 \{L_{ij},L_{ik},L_{jk}\}-(12+16a_k)L_{ij}^2-(12+16a_i)L_{jk}^2-(12+16a_j)L_{ik}^2\nonumber\\
\phantom{R_\ell^2=}{} +\frac{52}{3}(\{L_{ij},L_{ik}+L_{jk}\}+\{L_{ik},L_{jk}\})+
\left(\frac{16}{3}+\frac{176}{3}a_k\right)L_{ij}+\left(\frac{16}{3}+\frac{176}{3}a_i\right)L_{jk} \nonumber\\
\phantom{R_\ell^2=}{}
+\left(\frac{16}{3}+\frac{176}{3}a_j\right)L_{ij}
+64a_ia_ja_k +48(a_ia_j+a_ja_k+a_ka_i)+\frac{32}{3}(a_i+a_j+a_k),%\label{q6ordsym1}
\\
\frac{\epsilon_{ik\ell}\epsilon_{jk\ell}}2\{R_i,R_j\}
=\frac43(\{L_{i\ell},L_{jk},L_{k\ell}\}+\{L_{ik}, L_{j\ell}, L_{k\ell}\}-\{L_{ij}, L_{k\ell}, L_{k\ell}\}) +\frac{26}{3}\{L_{ik},L_{j\ell}\}\nonumber\\
\hphantom{\frac{\epsilon_{ik\ell}\epsilon_{jk\ell}}2\{R_i,R_j\}=}{}
+\frac{26}{3}\{L_{i\ell},L_{jk}\}+\frac{44}{3}\{L_{ij},L_{k\ell}\}+4L_{k\ell}^2-2\{L_{j\ell}+L_{jk}+L_{i\ell}+L_{ik},L_{k\ell}\}
\nonumber\\
\hphantom{\frac{\epsilon_{ik\ell}\epsilon_{jk\ell}}2\{R_i,R_j\}=}{}
-(6+8a_\ell)\{L_{ik},L_{jk}\}-(6+8a_k)\{L_{i\ell},L_{j\ell}\} -\frac{32}{3}L_{k\ell}\nonumber\\
\hphantom{\frac{\epsilon_{ik\ell}\epsilon_{jk\ell}}2\{R_i,R_j\}=}{}
-\left(\frac83-8a_\ell\right)(L_{jk}+L_{ik})-\left(\frac83-8a_k\right)(L_{jl}+L_{i\ell}) \nonumber\\
\hphantom{\frac{\epsilon_{ik\ell}\epsilon_{jk\ell}}2\{R_i,R_j\}=}{}
 +\left(\frac{16}{3}+24a_k+24a_\ell+32a_ka_\ell\right)L_{ij}
-16(a_ka_\ell+a_k+a_\ell).%\label{q6ordsym2}
\end{gather*}
Here $\{A,B,C\}=ABC+ACB+BAC+BCA+CAB+CBA$.

The eighth order functional relation is
\begin{gather*}
\sum_{i,j,k,l}\left[ \frac1{8} L_{ij}^2 L_{kl}^2
-\frac1{92} \lbrace L_{ik}, L_{il}, L_{jk}, L_{jl}\rbrace
-\frac{1}{36} \lbrace L_{ij}, L_{ik}, L_{kl}\rbrace  -\frac{7}{62} \lbrace L_{ij},L_{ij},L_{kl}\rbrace \right.\nonumber\\
\qquad {} + \frac1{6}\left(\frac12+\frac23 a_l\right) \lbrace L_{ij}L_{ik}L_{jk}\rbrace
+\frac 2 3  L_{ij}L_{kl}
-\left(\frac13-\frac{3}{4}a_k-\frac{3}{4} a_l-a_ka_l\right)L_{ij}^2\nonumber\\
\qquad{}
+\left(\frac 13 +\frac16 a_l\right) \lbrace L_{ik}, L_{jk} \rbrace
+\left(\frac{4} 3 a_k+\frac{4}3 a_l +\frac{7}{3}a_ka_l\right)L_{ij}\nonumber\\
\left.\qquad{}
+\frac{2}{3}a_{i}a_{j}a_{k}a_{l}+2a_{i}a_{j}a_{k}+\frac{4}{3}a_{i}a_{j}\right]=0.%\label{q8ordsym}
\end{gather*}
Here $\{A,B,C,D\}$ is the 24 term symmetrizer of 4 operators and the sum is taken over all pairwise distinct $i$, $j$, $k$, $\ell$.
For the purposes of the representation, it is useful to redef\/ine the constants as $a_i=b_i^2-\frac14. $

We note that the algebra described above contains several copies of the algebra generated by the corresponding potential
 on the two-sphere. Namely, let us def\/ine ${\cal A}$ to be the algebra generated by the set $\{L_{ij}, { I} \}$ for all $i,j=1,\dots,4$
where $I$ is the identity operator. Then, we can see that there exist subalgebras ${\cal A}_{k}$ generated by the set $\{L_{ij}, { I} \}$ for $i,j
\ne k $ and that these algebras are exactly those  associated to the 2D analog of this system. Furthermore, if we def\/ine
\[
H_k\equiv \sum_{i<j, i,j\ne k}L_{ij}-\left(\sum_{j\ne k}b_{i}^2 -\frac34\right){ I}
\]
then $H_k$ will commute with all the elements of ${\cal A}_k$ and will represent the Hamiltonian for the associated system. For example, take ${\cal A}_4$
to be the algebra generated by the set $\lbrace L_{12}, L_{13}, L_{23}, { I} \rbrace$. In this algebra, we have the
operator $H_4=L_{12}+L_{13}+L_{23}+(3/4-b_1^2-b_2^2-b_3^2){ I}$ which is in the center of ${\cal A}_4$ and which is the Hamiltonian
for the associated system on the two sphere immersed in $\mathbb{R}^3=\{(x_1,x_2,x_3)\}$.

Next we construct families of f\/inite dimensional and inf\/inite dimensional bounded below irreducible representations of this
algebra that include those that arise from the bound states of the associated quantum mechanical eigenvalue problem. At the same time we will
construct models of these representations via divided dif\/ference operators in two variables $s$ and $t$. Important tools for this construction are the
results of~\cite{KMP2007} giving the representations of the~${\cal A}_k$'s and known recurrence relations for one-variable
Wilson and Racah polynomials.

\section{Review of Wilson polynomials}
Before we proceed to the model, we us present a basic overview of some of the characteristics of the Wilson polynomials \cite{Wilson1980}
 that we plan to employ in the creation of our model. The polynomials are given by the expression
\begin{gather*}
w_n\left(t^2\right)\equiv w_n\left(t^2,\alpha,\beta,\gamma,\delta\right)=(\alpha+\beta)_n(\alpha+\gamma)_n(\alpha+\delta)_n\nonumber\\
\phantom{w_n(t^2)\equiv}{} \times
  {}_4F_{3}\left(\begin{array} {llll}-n,&\alpha+\beta+\gamma+\delta+n-1,&\alpha-t,&\alpha+t \\ \alpha+\beta,&\alpha+\gamma,&\alpha+\delta\end{array};1\right)\nonumber\\
\phantom{w_n(t^2)}{}
 = (\alpha+\beta)_n(\alpha+\gamma)_n(\alpha+\delta)_n\Phi^{(\alpha,\beta,\gamma,\delta)}_{n}\left(t^2\right),%\label{Wilson}
\end{gather*}
where $(a)_n$ is the Pochhammer symbol and ${}_4F_3(1)$ is a generalized hypergeometric function of unit argument. The polynomial $w_n(t^2)$ is symmetric in $\alpha$, $\beta$, $\gamma$, $\delta$.

The Wilson polynomials are eigenfunctions of a divided dif\/ference operator given as
\begin{gather} \label{wilsoneigen}
\tau^*\tau \Phi_n=n(n+\alpha+\beta+\gamma+\delta-1)\Phi_n ,
\end{gather}
where
\begin{gather*}
E^AF(t)=F(t+A),\qquad
\tau=\frac{1}{2t}\big(E^{1/2}-E^{-1/2}\big),\\
  \tau^*=\frac{1}{2t}\big[(\alpha+t)(\beta+t)(\gamma+t)(\delta+t)E^{1/2}-(\alpha-t)(\beta-t)(\gamma-t)(\delta-t)E^{-1/2}\big].
\end{gather*}
See \cite{MIL1987} for a simple derivation.

The Wilson polynomials $\Phi_n(t^2)\equiv\Phi^{(\alpha,\beta,\gamma,\delta)}_{n}(t^2)$, satisfy the three term recurrence formula
\begin{gather*}%\label{Wilsonrecurrence}
t^2\Phi_n\left(t^2\right)=K(n+1,n)\Phi_{n+1}\left(t^2\right)+K(n,n)
\Phi_n\left(t^2\right)+K(n-1,n)\Phi_{n-1}\left(t^2\right),
\end{gather*}
where
\begin{gather}
K(n+1,n)=\frac{\alpha+\beta+\gamma+\delta+n-1}{ (\alpha+\beta+\gamma+\delta+2n-1)(\alpha+\beta+\gamma+\delta+2n)}\nonumber\\
 \phantom{K(n+1,n)=}{} \times (\alpha+\beta+n)(\alpha+\gamma+n)(\alpha+\delta+n),\label{KWilson1} \\
 \label{KWilson2}
 K(n-1,n)=\frac{n (\beta+\gamma+n-1)(\beta+\delta+n-1)(\gamma+\delta+n-1) }{(\alpha+\beta+\gamma+\delta+2n-2)(\alpha+\beta+\gamma+\delta+2n-1)},\\
\label{KWilson3} K(n,n)=\alpha^2-K(n+1,n) -K(n-1,n).
\end{gather}
This formula, together with $\Phi_{-1}=0$, $\Phi_0=1$, determines the polynomials uniquely.

We can construct other recurrence relations between Wilson polynomials of dif\/ferent parameters using a family of divided dif\/ference operators $L_{\mu, \nu} $, $R^{\mu, \nu}$, $\mu, \nu =\alpha, \beta, \gamma, \delta$ given in Appendix~\ref{AppendixA}. Most importantly for the model considered below, we can construct operators which f\/ix $n$, the degree of the polynomial and which change the parameters by integer values. In the model constructed below, we will want to change $\alpha$ and $\delta$ by integer values and keep~$\beta$,~$\gamma$ f\/ixed. The operators which accomplish this are given by
\begin{gather*} L_{\alpha\beta}L_{\alpha\gamma} \Phi_n^{(\alpha,\beta,\gamma,\delta)}=(\alpha+\beta-1)(\alpha+\gamma-1)\Phi_n^{(\alpha-1,\beta,\gamma,\delta+1)},\\
R^{\alpha\beta}R^{\alpha\gamma} \Phi_n^{(\alpha,\beta,\gamma,\delta)}=
\frac{(n+\alpha+\beta)(n+\alpha+\gamma)(n+\beta+\delta-1)(n+\gamma+\delta-1)}{(\alpha+\beta)(\alpha+\gamma)}\Phi_n^{(\alpha+1,
\beta,\gamma,\delta-1)}.
\end{gather*}
We give the action on the $\Phi^{(\alpha,\beta,\gamma,\delta)}_{n}(t^2)$ for simplicity. For a complete exposition on the recurrence relations see Appendix~ \ref{AppendixA}.

Finally, the weight function of the model will be based on a two dimensional generalization of the weight function of the Wilson polynomials.

For f\/ixed $\alpha,\beta,\gamma,\delta>0$ (or if they occur in complex conjugate pairs with positive real parts)~\cite{Wilson1980}, the Wilson polynomials are orthogonal with respect to the inner product
\begin{gather}
\langle w_n,w_{n'}\rangle =\frac{1}{2\pi}\int_0^\infty w_n\left(-t^2\right)w_{n'}\left(-t^2\right) \left|\frac{\Gamma(\alpha+it)\Gamma(\beta+it)\Gamma(\gamma+it)\Gamma(\delta+it)}{\Gamma (2it)}\right|^2\ dt\nonumber\\
\phantom{\langle w_n,w_{n'}\rangle}{}
=\delta_{nn'}n!(\alpha+\beta+\gamma+\delta+n-1)_n\nonumber\\
\phantom{\langle w_n,w_{n'}\rangle=}{}
\times\frac{\Gamma(\alpha\!+\!\beta\!+\!n)\Gamma(\alpha\!+\!\gamma\!+\!n)\Gamma(\alpha\!+\!\delta\!+\!n)
\Gamma(\beta\!+\!\gamma\!+\!n)\Gamma(\beta\!+\!\delta\!+\!n)\Gamma(\gamma\!+\!\delta\!+\!n)}{\Gamma(\alpha\!+\!\beta\!+\!\gamma\!+\!\delta\!+\!2n)}.\!\!
\label{Wilsonnorm}
\end{gather}

When $m$ is a nonnegative integer then $\alpha+\beta=-m<0$ so that the above continuous Wilson orthogonality does not apply. The representation becomes f\/inite dimensional and the orthogonality is a f\/inite sum
\begin{gather}
\langle w_n,w_{n'}\rangle = \frac{(\alpha-\gamma+1)_m(\alpha-\delta+1)_m}{(2\alpha+1)_m(1-\gamma-\delta)_m}
\sum_{k=0}^m\frac{(2\alpha)_k(\alpha+1)_k(\alpha+\beta)_k(\alpha+\gamma)_k(\alpha+\delta)_k}{(1)_k(\alpha)_k(\alpha-\beta+1)_k
(\alpha-\gamma+1)_k(\alpha-\delta+1)_k}\nonumber\\
\phantom{\langle w_n,w_{n'}\rangle =}{}
\times w_n((\alpha+k)^2)w_{n'}((\alpha+k)^2)=\delta_{n{n'}}\nonumber\\
\phantom{\langle w_n,w_{n'}\rangle =}{}
\times \frac{n!(n\!+\!\alpha\!+\!\beta\!+\!\gamma\!+\!\delta\!-\!1)_n(\alpha\!+\!\beta)_n(\alpha\!+\!\gamma)_n(\alpha\!+\!\delta)_n(\beta\!+\!\gamma)_n
(\beta\!+\!\delta)_n(\gamma\!+\!\delta)_n}{(\alpha\!+\!\beta\!+\!\gamma\!+\!\delta)_{2n}}.\!\!\!
\label{Racah}
\end{gather}
Thus, the spectrum of the multiplication operator  $t^2$ is the set $\{ (\alpha+k)^2:\ k=0,\dots, m\}$.
Now, we are ready to determine the model.

\section{Construction of the operators for the model}

To begin, we review some basic facts about the representation.

The original quantum  spectral problem for (\ref{qham}) was studied in \cite{KMT} from an
entirely dif\/ferent point of view. It follows from this  study  that for the f\/inite dimensional irreducible
 representations of the quadratic algebra the multiplicity of each energy eigenspace is $(M+2)(M+1)/2$ and  we have
\begin{gather}\label{angmom}
L_{12}+L_{13}+L_{23}+L_{14}+L_{24}+L_{34}=\left(-\left(2M+\sum_{j=1}^4 b_j+3\right)^2 +\sum_{j=1}^4 b_j^2\right) I,
\end{gather}
where $I$ is the identity operator.

Of course, for an irreducible representation, the Hamiltonian will have to be represented by a constant times the identity and
initially for the construction of the model, we assume
\[
L_{12}+L_{13}+L_{23}+L_{14}+L_{24}+L_{34}=\left(E-1+\sum_{j=1}^4 b_j^2\right)I.
\]
We will obtain the quantized values of $E$ from the model.

We recall that each operator $L_{ij}$ is a member of the subalgebras ${\cal A}_{k}$ for  $k\ne i,j$. Thus, we can use the
known representations of these algebras, and symmetry in the indices,  to see that the eigenvalues
of each operator will be associated with eigenfunctions $\phi_{h,m}$ indexed by integers $0\leq h\leq m$ so that
\begin{gather} \label{eigenLij}
(L_{ij}) \phi_{h, m}=   \left(-(2h+b_i+b_j+1)^2-\frac12+b_i^2+b_j^2\right)\phi_{h ,m}, \\
\label{eigenHij}
(L_{ij}+L_{ik}+L_{jk}) \phi_{h,m} =\left(\frac14-(2m+b_i+b_j+b_k+2)^2\right)\phi_{h,m}.
\end{gather}

\subsection[A basis for $L_{13}$, $L_{12}+L_{13}+L_{23}$]{A basis for $\boldsymbol{L_{13}}$, $\boldsymbol{L_{12}+L_{13}+L_{23}}$}

As described above,  we seek to construct a representation of  ${\cal A}$ by extending the representations obtained for the
subalgebras ${\cal A}_k$.  The most important dif\/ference for our new representation is that the operator
$H_4=L_{12}+L_{13}+L_{23}+3/4-(b_1^2+b_2^2+b_3^2)$ is in the center of ${\cal A}_4$ but not~$\cal A$. Hence, it can no
longer be represented as a constant. We can still use the information about its eigenvalues to make an informed choice for
its realization.

 Restricting to bounded below irreducible representations of the quadratic algebra initially, we see from the
representations of ${\cal A}_4$ that the possible eigenvalues of $H_4$ are given as in~(\ref{eigenHij}) and the eigenvalues of
$L_{13}$ are given as in~(\ref{eigenLij}).

We can begin our construction of a two-variable model for the realization of these representations by choosing variables  $t$ and $s$, such that
\begin{gather*}
%\label{model1a}
H_4=\frac14-4s^2,\qquad L_{13}=-4t^2-\frac12+b_1^2+b_3^2,
\end{gather*}
i.e., the action of these operators is multiplication by the associated transform variables.  From the eigenvalues of the operators, we can see that the spectrum of $s^2$ is  $\{(-s_m)^2= (m+1+(b_1+b_2+b_3)/2)^2 \}$ and the spectrum of $t^2$ is $\{ t_\ell^2=(\ell+(b_1+b_2+1)/2)^2\} $.

In this basis, the eigenfunctions $d_{\ell,m}$ for a f\/inite dimensional representation are given by delta functions
\begin{gather*}%\label{dbasis1}
d_{\ell ,m}(s,t)=\delta(t-t_\ell)\delta(s-s_m),\qquad 0\le \ell \le m \le M.
\end{gather*}

\subsection[A basis for $L_{12}$, $L_{12}+L_{13}+L_{23}$]{A basis for $\boldsymbol{L_{12}}$, $\boldsymbol{L_{12}+L_{13}+L_{23}}$}\label{section4.2}

Next, we  construct $L_{12}$ in the model. Let $f_{n,m}$ be a basis for the model corresponding to simultaneous eigenvalues of
  $L_{12}$, $L_{12}+L_{13}+L_{23}$. From the representations of ${\cal A}_4$~\cite{KMP2007}, we know that the action of
 $L_{13}$ on this basis is given by
\begin{gather}\label{L23action}
L_{13}f_{n,m}=\sum_{j=n,n\pm 1}C_m(j,n)f_{j,m},
\end{gather}
where
\begin{gather} C_m(n,n) = \frac12\frac{(b_1^2-b_2^2)(b_1+b_2+2m+2)(b_1+b_2+2b_3+2m+2)}{(2n+b_1+b_2+2)(2n+b_1+b_2)}
  +b_1^2+b_3^2,\label{cnnident1} \\
 C_m(n,n+1)C_m(n+1,n)= 16(n+1)(n-m)(n-b_3-m)(n+b_2+1)(n+b_1+1)\nonumber\\
\qquad{} \times(n+b_1+b_2+1)
\frac{(n+m+b_1+b_2+2)(n+m+b_1+b_2+b_3+2)}{(2n+b_1+b_2+3)(2n+b_1+b_2+2)^2(2n+b_1+b_2+1)}. \label{cnn+1ident1}
\end{gather}

We already know that the bounded below representations of ${\cal A}_4$  are intimately connected with the Wilson polynomials.
The connection between these polynomials and the representation theory is the three term recurrence formula~(\ref{L23action})
for the action of~$L_{13}$ on an~$L_{12}$ basis, where the coef\/f\/icients are given by~(\ref{cnnident1}) and~(\ref{cnn+1ident1}).

We def\/ine the operator $L$ on the representation space of the superintegrable system by the action of the three term recurrence relations
for the Wilson polynomials given by expansion coef\/f\/icients (\ref{KWilson1})--(\ref{KWilson3}),  i.e.
\begin{gather*}%\label{L4def}
Lf_n=K(n+1,n)f_{n+1}+K(n,n)f_n+K(n-1,n)f_{n-1}.
\end{gather*}
Note that with the choices
\begin{alignat}{3}
& \alpha =   -\frac{b_1+b_3+1}{2}-m,\qquad &&  \beta=\frac{b_1+b_3+1}{2}, &\nonumber \\
 & \gamma= \frac{b_1-b_3+1}{2},\qquad &&  \delta=\frac{b_1+b_3-1}{2}+b_2+m+2,& \label{param}
 \end{alignat}
we have a perfect match with the action of $L_{13}$ as
\[
 C_m(n+1,n)=4K(n+1,n),\qquad C_m(n-1,n)=4K(n-1,n)-\frac12+b_1^2+b_3^2.
\]
Thus, the action of $L_{13}$ is given by
\[
L_{13}f_n=\left(-4L-\frac12+b_1^2+b_3^2\right)f_n,
\]
and so we see  that the action of $L_{13}$ on an $L_{12}$ basis is exactly the action of the variable $t^2$ on a basis of Wilson polynomials.
 Hence, we hypothesize that $L_{12}$ takes the form of an eigenvalue operator for Wilson polynomials in the variable $t$
\begin{gather*}%\label{model1b}
L_{12}=-4\tau_t^*\tau_t-2(b_1+1)(b_2+1)+1/2,
\end{gather*}
where $\tau$, $\tau_t^*$ are given as (\ref{wilsoneigen}) with the choice of parameters as given in (\ref{param}). Here the subscript~$t$ expresses the fact that this is a  dif\/ference operator in the variable~$t$, although the parameters depend on the variable~$s$.

The basis functions corresponding to diagonalizing $H_4$ and $L_{12}$ can be taken, essentially,  as  the Wilson polynomials
\begin{gather*}
f_{n,m}(t,s)=w_n\left(t^2,\alpha,\beta,\gamma,\delta\right)\delta(s-s_m),
\end{gather*}
where $s_m=m+1+(b_1+b_2+b_3)/2$ as above.  Note that $w_n(t^2)$ actually depends on $m$ (or $s^2$) through
the parameters~$\alpha$,~$\delta$. Also $\alpha+\delta $ is independent of~$m$.
Written in terms of the variable~$s$, the parameters are given by
\begin{gather} \label{para1}
\alpha=\frac{b_2+1}{2}+s,\qquad  \beta =\frac{b_1+b_3+1}{2},\qquad \gamma=\frac{b_1-b_3+1}{2},\qquad \delta=\frac{b_2+1}{2}-s.
\end{gather}
Note that when $s$ is restricted to $s_m$, these parameters agree with (\ref{param}).

Since the $w_n$ are symmetric with respect to arbitrary permutations of  $\alpha$, $\beta$, $\gamma$, $\delta$, we can transpose~$\alpha$ and~$\beta$ and verify that~$w_n$ is a polynomial of order~$n$ in~$s^2$.

\subsection[A basis for $L_{13}$, $L_{24}$]{A basis for $\boldsymbol{L_{13}}$, $\boldsymbol{L_{24}}$}

For now, let us assume that we have a f\/inite dimensional irreducible representation such that the simultaneous eigenspaces of $L_{12}$, $L_{12}+L_{13}+L_{23}$ are indexed by integers $n$, $m$, respectively, such that $0\le n\le m \le M$.
Each simultaneous eigenspace is one-dimensional and the total dimension of the representation space is $(M+1)(M+2)/2$.
Now we need to determine the action of the operators $L_{14}$, $L_{24}$, $L_{34}$ in the model.

 A reasonable guess of the form of the operator $L_{24}$ is as a dif\/ference operator in $s, $ since it commutes with $L_{13}$.
We hypothesize that it takes the form of an eigenvalue equation for the Wilson polynomials in the variable $s$. We require that it have eigenvalues of the form~(\ref{eigenLij}).

Note that when acting on the delta basis $d_{\ell,m}$, it produces a three-term recursion relation. For our representation, we require that that the representation cut of\/f at the appropriate bounds. That is if we write the expansion coef\/f\/icients of $L_{24}$ acting on $d_{\ell,m}$, as
\[
L_{24}d_{\ell,m}=B(m,m-1)d_{\ell,m-1}+B(m, m)d_{\ell,m}+B(m, m+1)d_{\ell,m+1} ,
\]
we require $B(m, m-1)B(m-1, m)=0$ and $B(M, M+1)B(M+1, M)=0$.
These restrictions are realized in our choices of parameters,
\begin{alignat}{3}
&  {\tilde \alpha} = t+\frac{b_2+1}{2},\qquad &&  {\tilde \beta}=-M-\frac{b_1+b_2+b_3}{2}-1,& \nonumber\\
& {\tilde \gamma} = M+b_4+\frac{b_1+b_2+b_3}{2}+2,\qquad && {\tilde \delta}=-t+\frac{b_2+1}{2}. & \label{tildeparameters}
 \end{alignat}
For $L_{24}$ we take
\begin{gather*}%\label{model1c}
L_{24}=-4{\tilde \tau}^*_s{\tilde \tau}_s-2(b_2+1)(b_4+1)+\frac12.
\end{gather*}
Here ${\tilde \tau}_s$ is the dif\/ference operator in $s$ where the parameters are ${\tilde \alpha}$, ${\tilde \beta}$,
${\tilde \gamma}$, ${\tilde \delta}$.

With the operator $L_{24}$ thus def\/ined, the unnormalized eigenfunctions of the commuting operators $L_{13}$, $L_{24}$ in the model take the form
$g_{n,k}$ where
\begin{gather*}
 L_{13}g_{\ell,k}=\left(-(2\ell+b_1+b_3+1)^2-\frac12+b_1^2+b_3^2\right)g_{n,k},\\
  L_{24}g_{\ell.k}=\left(-(2k+b_2+b_4+1)^2-\frac12+b_2^2+b_4^2\right)g_{n,k},
\end{gather*}
where $0\le \ell \le M$, $0\le k\le M-\ell$, and
\begin{gather}\label{L13L24basis}
g_{\ell,k}=\delta(t-t_\ell)w_k\big(s^2,{\tilde \alpha},{\tilde \beta},{\tilde \gamma},{\tilde \delta}\big),
\end{gather}
with $t_\ell=\ell+(b_1+b_3+1)/2$ as above.

For this choice of parameters, the functions (\ref{L13L24basis}) constitute an alternative basis for the representation space, consisting of polynomials in $s^2$, $t^2$ multiplied by a delta function in $s$.

\subsection{Completion of the model}

In this section, we f\/inalize the construction of our model by realizing the operator $L_{34}$.
The operator $L_{34}$ must commute with $L_{12}$, so we hypothesize that it is of the form
\begin{gather}\label{model1d}
L_{34}=A(s)S(L_{\alpha\beta}L_{\alpha\gamma})_t+B(s)S^{-1}(R^{\alpha\beta}R^{\alpha\gamma})_t+C(s)(LR)_t+D(s),
\end{gather}
where $S^uf(s,t)=f(s+u,t)$,  $A$, $B$, $C$, $D$ are rational functions of $s$ to be determined, and the operators $L_{\alpha\beta}$, $R^{\alpha\beta}$, $L$, $R$, etc.\ are def\/ined in Appendix~\ref{AppendixB}.  The subscript $t$ denotes  dif\/ference operators in~$t$.
 (Note that $\tau^*_t\tau\equiv (LR)_t$.)  The parameters are (\ref{para1}).
Here
\begin{gather*} L_{\alpha\beta}L_{\alpha\gamma}=
 \frac{1}{4t(t+\frac12)}(\alpha-1+t)(\alpha+t)(\beta+t)(\gamma+t)T^1\nonumber\\
 \phantom{L_{\alpha\beta}L_{\alpha\gamma}= }{}
 +\frac{1}{4t(t-\frac12)}(\alpha-1-t)(\alpha-t)(\beta-t)
(\gamma-t)T^{-1}
\nonumber\\
 \phantom{L_{\alpha\beta}L_{\alpha\gamma}= }{}
-\frac{1}{4t(t+\frac12)}(\alpha-1+t)(\alpha-1-t)(\beta+t)(\gamma-1-t)
\nonumber\\
 \phantom{L_{\alpha\beta}L_{\alpha\gamma}= }{}
-\frac{1}{4t(t-\frac12)}(\alpha-1-t)(\alpha-1+t)(\beta-t)
(\gamma-1+t),\\ %\label{LabLag}\\
 R^{\alpha\beta}R^{\alpha\gamma}=
 \frac{1}{4t(t+\frac12)}(\beta+t)(\gamma+t)(\delta-1+t)(\delta+t)T^1\nonumber\\
\phantom{R^{\alpha\beta}R^{\alpha\gamma}=}{}
 +\frac{1}{4t(t-\frac12)}(\beta-t)(\gamma-t)
(\delta-1-t)(\delta-t)T^{-1}
\nonumber\\
\phantom{R^{\alpha\beta}R^{\alpha\gamma}=}{}
-\frac{1}{4t(t+\frac12)}(\beta-1-t)(\gamma+t)(\delta-1+t)(\delta-1-t)
\nonumber\\
\phantom{R^{\alpha\beta}R^{\alpha\gamma}=}{}
-\frac{1}{4t(t-\frac12)}(\beta-1+t)(\gamma-t)
(\delta-1-t)(\delta-1+t), \\ %\label{RabRag}\\
 LR= \frac{1}{4t(t+\frac12)}(\alpha+t)(\beta+t)(\gamma+t)(\delta+t)T^1+\frac{1}{4t(t-\frac12)}(\alpha-t)(\beta-t)(\gamma-t)
(\delta-t)T^{-1} \nonumber\\
\phantom{LR=}{}
-\frac{1}{4t(t + \frac12)}(\alpha+t)(\beta+t)(\gamma+t)(\delta+t)-\frac{1}{4t(t - \frac12)}(\alpha-t)(\beta-t)(\gamma-t)(\delta-t).%\!\!\!\! \label{LRabgd}
\end{gather*}

On the other hand, we can consider the action of $L_{34}$ on the basis (\ref{L13L24basis}). Considering $L_{34}$ primarily as an operator on $s$ we hypothesize that it must be of the form
\begin{gather}\label{model1dd}L_{34}={\tilde A}(t)T(L_{{\tilde \alpha}{\tilde\beta}}L_{{\tilde \alpha}{\tilde\gamma}})_s+{\tilde B}(t)T^{-1}(R^{{\tilde \alpha}{\tilde\beta}}R^{{\tilde\alpha}{\tilde\gamma}})_s+{\tilde C}(t)(LR)_s
 +{\tilde D(t)}s^2+{\tilde E}(t)+\kappa L_{12},\!\!\!
  \end{gather}
where the dif\/ference operators are def\/ined in Appendix~\ref{AppendixB} with subscript $s$ denoting
  dif\/ference operators in $s$ and $\kappa$ is a constant.

Finally, we express the operator $L_{14}$ as
\[
\left(E+\sum_{j=1}^4 b_j^2-1\right)I-L_{12}-L_{13}-L_{23}-L_{24}-L_{34}.
\]

By a long and tedious computation we can verify that the 3rd order structure equations are satisf\/ied  if and only if $E$ takes the values
 \[
 E=-\left(2M+\sum_{j=1}^4 b_j+3\right)^2-1
 \]
  and  the  functional coef\/f\/icients for $L_{34}$ in   (\ref{model1d}), (\ref{model1dd})  take the following form :
\begin{gather}
A(s)=-\frac{(2 M+b_1+b_2+b_3-2 s+2) (2 M+b_1+b_2+b_3+2 b_4+2 s+4)}{2s (2 s+1)},\nonumber\\
        B(s)=-\frac{(2M+b_1+b_2+b_3+2 s+2) (2 M+b_1+b_2+b_3+2 b_4-2 s+4)}{2s (2 s-1)},\nonumber\\
        C(s)=-2+\frac{2(2M+b_1+b_2+b_3+3)(2M+b_1+b_2+b_3+2b_4+3)}{4s^2-1},\nonumber\\
        D(s) = 2s^2-2\left(\frac{2M+b_1+b_2+b_3+b_4+4}{2}\right)^2\!\!-\frac{(b_1+b_2)^2+b_3^2+b_4^2}{2}+b_3+b_4+2M+3\nonumber\\
      %\label{Ds}
      \phantom{D(s) =}{}
      +\frac{((b_1+b_2+1)^2-b_3^2)(2M+b_1+b_2+b_3+3)(2M+b_1+b_2+b_3+2b_4+3)}{2(4s^2-1)},\nonumber\\
      {\tilde A}(t) = \frac{(b_1-b_3+2 t+1) (b_1+1+b_3+2 t)}{2t (2 t+1)},\nonumber\\
      {\tilde B}(t) = \frac{(b_1-b_3-2 t+1) (b_1+1+b_3-2 t)}{2t (2 t-1)},\nonumber\\
      {\tilde C}(t) = 2+\frac{2(b_3^2-b_1^2)}{4 t^2-1},\nonumber\\
      {\tilde D}(t) = 2,\label{As}
      \end{gather}
  and $\kappa=-4$. The  expression for ${\tilde E}(t)$ takes the form  ${\tilde E}(t)=\mu_1+\mu_2/(4t^2-1)$ where
 $ \mu_1$, $\mu_2$ are constants, but  we will not  list it  here in detail.

For f\/inite dimensional representations, we have the requirement that $M$ be a positive integer so we obtain the quantization of
the energy obtained previously~(\ref{angmom}).

\subsection{The model and basis functions}
We shall now review what we have constructed, up to this point. We realize the algebra $\cal A$ by the following operators
\begin{gather*}
   H= -\left(\left(2M+\sum_{j=1}^4 b_j+3\right)^2+1\right)I, \\
 H_4 = \frac14-4s^2,\qquad
  L_{13} = -4t^2-\frac12+b_1^2+b_3^2,\nonumber\\
  L_{12} = -4\tau_t^*\tau_t-2(b_1+1)(b_2+1)+\frac12,\qquad
  L_{24} =  -4{\tilde \tau}^*_s{\tilde \tau}_s-2(b_2+1)(b_4+1)+\frac12,\nonumber\\
  L_{34} =  A(s)S(L_{\alpha\beta}L_{\alpha\gamma})_t+B(s)S^{-1}(R^{\alpha\beta}R^{\alpha\gamma})_t+C(s)(LR)_t+D(s),\nonumber
  \end{gather*}
  where the parameters for the $\tau_t$ operators are given in~(\ref{para1}), the parameters for the operators~${\tilde \tau}_s$ are given in~(\ref{tildeparameters}) and the functional coef\/f\/icients of~$L_{34}$ are given in~(\ref{As}). %-\ref{Ds}).
  The operators $L_{23}$, $L_{14}$ can be obtained through linear combinations of this basis.

Using Maple, we have verif\/ied explicitly that this solution satisf\/ies all of the 4th, 5th, 6th and 8th order structure equations.

We have computed three sets of orthogonal basis vectors corresponding to diagonalizing three sets of commuting operators,
 $\{ L_{13},H_4\}$, $\{L_{12}, H_4\}$ and $\{ L_{13}, L_{24}\}$, respectively,
\begin{gather} \label{dbasis}
{\tilde d}_{\ell,m}(s,t) = \delta(t-t_\ell )\delta(s-s_m),\qquad  0\le \ell \le m\le M,\\
\label{fbasis}
f_{n,m}(s,t) = w_n(t^2,\alpha,\beta,\gamma,\delta)\delta(s-s_m),\qquad 0\le n\le m\le M,\\
\label{gbasis}g_{\ell,k}(s,t) = w_k(s^2,\tilde{\alpha},\tilde{\beta},\tilde{\gamma},\tilde{\delta})\delta(t-t_\ell),\qquad  0\le \ell \le k+\ell \le M.
\end{gather}
We also have a nonorthogonal basis given by
\begin{gather*}%\label{hbasis}
h_{n,k}(s,t)=t^{2n}s^{2k},\qquad 0\le n+k\le M.
\end{gather*}
Recall that the spectrum of the variables $s$, $t$ is  given by
\begin{gather*}%\label{spectrum}
t_\ell=\ell+\frac{b_1+b_3+1}{2},\qquad s_m=-\left(m+1+\frac{b_1+b_2+b_3}{2}\right),\qquad 0\le \ell \le m\le M.
\end{gather*}

We f\/inish the construction of the model by computing normalizations for the basis $f_{n,m}$, and~$ g_{\ell, m}$  and the weight function.

\section{The weight function and normalizations}

We begin this section by determining the weight function and normalization of the basis functions in the f\/inite dimensional representations.
 Later, we shall extend the system to the inf\/inite dimensional bounded below case.

\subsection[The weight function and normalization of the basis $ {\tilde d}_{\ell,m}(s,t)=\delta(t-t_\ell )\delta(s-s_m)$]{The weight function and normalization\\ of the basis $\boldsymbol{{\tilde d}_{\ell,m}(s,t)=\delta(t-t_\ell )\delta(s-s_m)}$}

We consider the normalization for the $d_{\ell, m} =\delta(t-t_\ell)\delta(s-s_m)$ basis for f\/inite dimensional representations where
\[
t_\ell=\ell+\frac{b_1+b_3+1}{2},\qquad s_m=-\left(m+1+\frac{b_1+b_2+b_3}{2}\right),\qquad 0\le \ell \le m\le M.
\]
In order to derive these results we use the requirement that the generating operators $L_{ij}$ are formally self-adjoint.

Consider a weight function $\omega(t,s)$ so that
\begin{gather*} \langle f(t,s), g(t,s) \rangle=\iint f(t,s)g(t,s)\omega(t,s) dsdt,
\end{gather*}
then  we assume that the basis functions are orthonormal with
\begin{gather*}
\langle c_{\ell,m} \delta(t-t_\ell)\delta(s-s_m), c_{\ell',m}\delta(t-t_\ell')\delta(s-s_m')\rangle =\delta_{m,m'}\delta_{\ell,\ell'},
\end{gather*}
which implies that  $c^2_{\ell,m}\omega(t_\ell, s_m)=1$.
The adjoint properties of $L_{13}$ and $L_{24}$ provide recurrence relations on the $c_{\ell,m}$.
That is
\begin{gather*} \langle \delta(t-t_\ell-1)\delta(s-s_m), L_{13}\delta(t-t_\ell)\delta(s-s_m)\rangle \\
\qquad{}=\langle L_{13} \delta(t-t_\ell-1)\delta(s-s_m), \delta(t-t_\ell)\delta(s-s_m)\rangle
\end{gather*}
implies the recurrence relation
\begin{gather} \label{cn} \frac{c_{\ell+1,m}^2}{c^2_{\ell,m }}=\frac{(\ell+1)(1+b_3+\ell)(m-\ell+b_2)(m+\ell+b_1+b_3+2)(2\ell+b_1+b_3+1)}{(m-\ell)(1+b_1+b_3+\ell)
(m+\ell+b_1+b_2+b_3+2)(2\ell+b_1+b_3+3)}.
\end{gather}

Similarly,  the self-adjoint property of $L_{24}$
\begin{gather*} \langle \delta(t-t_\ell)\delta(s-s_m+1), L_{24}\delta(t-t_\ell)\delta(s-s_m)\rangle\\
\qquad{}=\langle L_{24} \delta(t-t_\ell-1)\delta(s-s_m+1), \delta(t-t_\ell)\delta(s-s_m)\rangle
\end{gather*}
implies the recurrence relation
\begin{gather} \frac{c^2_{\ell, m+1}}{c^2_{\ell, m}} = \frac{(M-\ell+b_4)(m+\ell+b_1+b_3+2)(M+m+b_1+b_2+b_3+2)}{(M+m+b_1+b_2+b_3+b_4+3)(m+\ell+b_1+b_2+b_3+2)}\nonumber\\
 \hphantom{\frac{c^2_{\ell, m+1}}{c^2_{\ell, m}} =}{}
  \times\frac{(m-\ell+1)(2m+2+b_1+b_2+b_3)}{(M-m)(m-\ell+1+b_2)(2m+4+b_1+b_2+b_3)}. \label{cm}
  \end{gather}
Putting together \eqref{cn} and \eqref{cm} we obtain
\begin{gather}
\frac{c_{\ell,m}^2}{c^2_{0,0}} = \frac{(1+b_3)_\ell(1+b_4)_M(M+b_1+b_2+b_3+3)_m(2+b_1+b_3)_{m+\ell}}{ (1+b_2)_{m-\ell}(1+b_1)_\ell(1+b_1+b_3)_\ell(1+b_4)_{M-m}
(M+b_1+b_2+b_3+b_4+3)_m}\nonumber\\
\hphantom{\frac{c_{\ell,m}^2}{c^2_{0,0}} =}{}  \times \frac{(M-m)!(m-\ell)!\ell!(2+b_1+b_2+b_3)(1+b_1+b_3)}{M!(2m+2+b_1+b_2+b_3)(2\ell+1+b_1+b_3)(2+b_1+b_2+b_3)_{m+\ell}},\label{cmnfinal}
\end{gather}
which gives the value of the weight function for  the spectrum of $t$, $s$ via $w(t_\ell, s_m)=c_{\ell,m}^{-2}$.

\subsection[Normalization of the $w_n(t^2)\delta(s-s_m)$ basis]{Normalization of the $\boldsymbol{w_n(t^2)\delta(s-s_m)}$ basis}

Next, we use the orthogonality of the Wilson polynomials to f\/ind the normalization of the $f_{n,m}$ basis in the
f\/inite dimensional representation.

Assume the normalized basis functions have the form
\begin{gather*}%\label{fbasis1}
\hat{f}_{n,m}(s,t)=k_{n,m} w_n\left(t^2,\alpha,\beta,\gamma,\delta\right)\delta(s-s_m),
\qquad 0\le n\le m\le M.
\end{gather*}

When evaluated at $s=s_m$, the parameters are given by
\begin{alignat}{3}
& \alpha =   -\frac{b_1+b_3+1}{2}-m, \qquad &&  \beta=\frac{b_1+b_3+1}{2}, & \nonumber\\
&  \gamma= \frac{b_1-b_3+1}{2} ,\qquad && \delta=\frac{b_1+b_3-1}{2}+b_2+m+2,&\label{param2}
\end{alignat}
and satisfy $\alpha+\beta=-m<0$. Thus, the Wilson orthogonality is realized as a f\/inite sum over the weights of $t^2$.
However, the weight of the variable $t$ is given by $t_\ell=\ell+\beta$ and we must adjust the equation for the Wilson
orthogonality~(\ref{Racah}) by permuting~$\alpha$ and~$\beta$. This is allowed since the polynomial and the requirement  $\alpha+\beta=-m$ are symmetric in the two parameters. In this form the Wilson orthogonality is given over the spectrum of the multiplication operator  $t^2$ as the set $\{ (\beta+\ell)^2:\ \ell=0,\dots, m\}$
\begin{gather}
\langle w_n,w_{n'}\rangle = \frac{(\beta-\gamma+1)_m(\beta-\delta+1)_m}{(2\beta+1)_m(1-\gamma-\delta)_m}  \label{Racahb}\\
  \times \sum_{\ell=0}^m\frac{(2\beta)_\ell(\beta+1)_\ell (\beta+\alpha)_\ell(\beta+\gamma)_\ell(\beta+\delta)_\ell}{(1)_\ell(\beta)_\ell(\beta-\alpha+1)_\ell(\beta-\gamma+1)_\ell(\beta-\delta+1)_\ell}  w_n((\beta+\ell)^2)w_{n'}((\beta+\ell)^2)\nonumber\\
 = \frac{\delta_{n{n'}}n!(n+\alpha+\beta+\gamma+\delta-1)_n(\alpha+\beta)_n(\beta+\gamma)_n(\beta+\delta)_n
 (\alpha+\gamma)_n(\alpha+\delta)_n(\gamma+\delta)_n}{(\alpha+\beta+\gamma+\delta)_{2n}}. \nonumber
\end{gather}

In light of this orthogonality, we hypothesize that the weight function is given by
\begin{gather*}
\langle f(t,s),g(t,s)\rangle =
\iint \sum_{\ell,m} f(t,s)g(t,s) w(t,s)\delta(t-t_\ell)\delta(s-s_m)\nonumber\\
\phantom{\langle f(t,s),g(t,s)\rangle}{}  =  \sum_{\ell,m}f(t_\ell,s_m)g(t_\ell,s_m)\omega(t_\ell,s_m)
\end{gather*}
and so we look for normalization constants so that
\begin{gather}  \langle \hat{f}_{n,m}(s,t) ,\hat{f}_{n',m'}(s,t)\rangle  = \iint \sum_{\ell,m}  \hat{f}_{n,m}(s,t)\hat{f}_{n',m'}(s,t)w(t,s)\delta(t-t_\ell)\delta(s-s_m)\nonumber\\
 \phantom{\langle \hat{f}_{n,m}(s,t) ,\hat{f}_{n',m'}(s,t)\rangle}{} = \delta_{m,m'} \sum_{\ell}k_{n,m}k_{n',m}w_n(t_\ell^2)w_{n'}(t_\ell^2) w(t_\ell,s_m)\delta(t-t_\ell)\delta(s-s_m)\nonumber\\
\phantom{\langle \hat{f}_{n,m}(s,t) ,\hat{f}_{n',m'}(s,t)\rangle}{}= \delta_{m, m'} \delta_{n,n'}.\label{innerf}
\end{gather}

The orthogonality (\ref{Racahb}) in terms of the choices of parameters (\ref{param2}) is given by
\begin{gather} \delta_{n, n'} =  \frac{(2+b_1+b_2)_{2n}(m-n)!(1+b_3)_{m-n}}
{(n +b_1+b_2+1)_n (1+b_1)_n(1+b_2)_n(2+b_1+b_2)_{m+n}(2+b_1+b_2+b_3)_{m+n}}\nonumber\\
\phantom{\delta_{n, n'} =}{} \times \sum_{\ell=0}^m\frac{(1+b_1)_\ell(1+b_1+b_3)_\ell(1+b_2)_{m-\ell}(2+b_1+b_2+b_3)_{m+\ell}}
{\ell!(m-\ell)!(1+b_3)_\ell(2+b_1+b_3)_{m+\ell}}\nonumber\\
%\label{wilsont}
\phantom{\delta_{n, n'} =}{}
\times \frac{(2\ell+b_1+b_3+1)}{(1+b_1+b_3)}w_n(t_\ell^2)w_{n'}(t_\ell^2).\label{racahell}
\end{gather}

The weight function \eqref{cmnfinal} can be rewritten as
\begin{gather}\label{weightt}
\omega(t_\ell, s_m)=\frac
{M!(1+b_4)_{M-m}(M+b_1+b_2+b_3+b_4+3)_m(2m+2+b_1+b_2+b_3)}
{(M-m)!(1+b_4)_M(M+b_1+b_2+b_3+3)_m(2+b_1+b_2+b_3)}\\
\phantom{\omega(t_\ell, s_m)=}{}  \times  \frac{(1+b_1)_\ell(1+b_1+b_3)_\ell(1+b_2)_{m-\ell}(2+b_1+b_2+b_3)_{m+\ell}(2\ell+1+b_1+b_3)}
{\ell!(m-\ell)!(1+b_3)_\ell(2+b_1+b_3)_{m+\ell}(1+b_1+b_3)c^2_{0,0}}.\nonumber
\end{gather}
We can now solve the equation (\ref{innerf}) for $k_{n,m}$ by comparing (\ref{weightt})  and (\ref{racahell}) to obtain
\begin{gather*}
\frac{k_{n,m}^2}{c_{0,0}^2} = \frac{(1+b_4)_M(1+b_3)_{m-n}(M+b_1+b_2+b_3+3)_m(2+b_1+b_2+b_3)(2+b_1+b_2)_{2n}}
{n!M!(1+b_4)_{M-m}(n +b_1+b_2+1)_n (1+b_1)_n(1+b_2)_n(2m+2+b_1+b_2+b_3)}\nonumber\\
\phantom{\frac{k_{n,m}^2}{c_{0,0}^2} =}{}
 \times\frac{(m-n)!(M-m)!}{(M+b_1+b_2+b_3+b_4+3)_m(2+b_1+b_2)_{m+n}(2+b_1+b_2+b_3)_{m+n}}.
\end{gather*}

With this normalization the basis functions $\hat{f}_{n,m}(s,t)$ are orthonormal.

\subsection[Normalization of the $w_k(s^2)\delta(t-t_\ell)$  basis]{Normalization of the $\boldsymbol{w_k(s^2)\delta(t-t_\ell)}$  basis}

Next, we use the orthogonality of the Wilson polynomials to f\/ind the normalization of the $g_{n,k}$ basis in the
f\/inite dimensional representation.
We take the normalized basis functions to be given by
\begin{gather*} %\label{ghat}
\hat{g}_{\ell,k}(s,t)=h_{\ell,k}w_k(s^2,\tilde{\alpha},\tilde{\beta},\tilde{\gamma},\tilde{\delta})\delta(t-t_\ell),\qquad 0\le n\le M,\qquad 0\le k\le M-\ell.\end{gather*}

Again, we want to show that there exist normalization constants $h_{\ell,k}$ so that the following holds:
\begin{gather*} \langle \hat{g}_{\ell,k}(s,t) ,\hat{g}_{\ell',k'}(s,t)\rangle =
\iint \sum_{\ell,m} \hat{g}_{\ell,k}(s,t)\hat{g}_{\ell',k'}(s,t)
w(t,s)\delta(t-t_\ell)\delta(s-s_m)\nonumber\\
\phantom{\langle \hat{g}_{\ell,k}(s,t) ,\hat{g}_{\ell',k'}(s,t)\rangle }{} = \delta_{\ell,\ell'} \sum_{m}h_{\ell,k}h_{\ell,k'}w_k(s_m^2)w_{k'}(s_m^2) w(t_\ell,s_m)\delta(t-t_\ell)\delta(s-s_m)\nonumber\\
\phantom{\langle \hat{g}_{\ell,k}(s,t) ,\hat{g}_{\ell',k'}(s,t)\rangle }{}
 =  \delta_{\ell, \ell'} \delta_{k,k'}.\end{gather*}

When restricted to $t=t_\ell$ the parameters $\tilde{\alpha}$, $\tilde{\beta}$, $\tilde{\gamma}$, $\tilde{\delta}$ become
\begin{alignat}{3}
& \tilde{\alpha} = \ell+1+\frac{b_1+b_2+b_3}{2},\qquad &&  \tilde{\beta}=-M-1-\frac{b_1+b_2+b_3}2,& \nonumber\\
& \tilde{\gamma}= M+b_4+2+\frac{b_1+b_2+b_3}{2},\qquad&&  \tilde{\delta}=-\ell-\frac{b_1-b_2+b_3}{2},& \label{tildeparam}
\end{alignat}
and so we have $\tilde{\alpha}+\tilde{\beta}=-M+\ell<0$. Also, the spectrum of the variable~$s^2$ is given by the set
 $\{ (m+1+\frac{b_1+b_2+b_3}{2})^2:\ m=\ell,\ldots, M\}$ which we can write as $\{ ((m-\ell)+\tilde{\alpha})^2 :\ m-\ell=0,\ldots, M-\ell\}$.

The Wilson orthogonality can be written in terms of the choice of parameters  (\ref{tildeparam})   as
\begin{gather*}  \delta_{k, k'} = \frac{(M-k+\ell)!(3+b_1+b_2+b_3)_{M}}{(3+b_1+b_2+b_3+b_4)_{M+k-\ell}}\nonumber\\
\phantom{\delta_{k, k'} =}{} \times \frac{(2+b_1+b_3)_{M+\ell-k}(3+b_1+b_2+b_3+b_4)_M(2+b_2+b_4+k)_{k}}
{(1+b_2)_{k}(1+b_4)_{k}(1+b_2+b_4+k)_{k}(2+b_2+b_4)_{M-\ell+k}}\nonumber\\
\phantom{\delta_{k, k'} =}{}\times \sum_{m=\ell}^{M}\frac{(1+b_4)_{M-m}(M+b_1+b_2+b_3+b_4+3)_m (2+b_1+b_2+b_3)_{m+\ell}}
{(M-m)!(m-\ell)!(M+b_1+b_2+b_3+3)_m(2+b_1+b_3)_{m+\ell}}\nonumber\\
\phantom{\delta_{k, k'} =}{} \times \frac{(1+b_2)_{m-\ell}(2m+2+b_1+b_2+b_3)}{(2+b_1+b_2+b_3)}w_{k}\left(s_m^2\right)
w_{k'}\left(s_m^2\right),%\label{wilsons}
\end{gather*}
where the index being summed over is $m=\ell, \ldots, M$ instead of $m-\ell=0,\ldots, M-\ell$.

Comparing this orthogonality with the weight function \eqref{cmnfinal} written as
\begin{gather*}
\omega(t_\ell, s_m) = \frac{(1+b_4)_{M-m}(M+b_1+b_2+b_3+b_4+3)_m(2+b_1+b_2+b_3)_{m+\ell}}
{(M-m)!(m-\ell)!(M+b_1+b_2+b_3)_m(2+b_1+b_3)_{m+\ell}}\nonumber\\
\phantom{\omega(t_\ell, s_m) =}{} \times\frac{(1+b_2)_{m-\ell}(2m+2+b_1+b_2+b_3) }{(2+b_1+b_2+b_3)}\nonumber\\
\phantom{\omega(t_\ell, s_m) =}{} \times \frac{M!(1+b_1)_\ell(1+b_1+b_3)_\ell(2\ell+1+b_1+b_3)}
{\ell!(1+b_3)_\ell(1+b_4)_M(2+b_1+b_2+b_3)(1+b_1+b_3)c^2_{0,0}},
\nonumber
\end{gather*}
 the normalization constants are determined by the requirement
\begin{gather*}
\delta_{k, k'}=h_{\ell, k }l_{\ell, k}\sum_{m=\ell}^M\omega(t_\ell, s_m)w_{k}\left(s_m^2\right)w_{k}\left(s_m^2\right)
\end{gather*}
for $l_{\ell, k}$.
The proper choice of normalization is
\begin{gather*}
\frac{h_{\ell, k} }{c^2_{0,0}} = \frac{(M-k+\ell)!(2+b_1+b_3)_{M+k+\ell}(2+b_2+b_4)_{2k}}{(1+b_2)_{k}(1+b_4)_{k}(1+b_2+b_4+k)_{k}
(3+b_1+b_2+b_3)_M(2+b_2+b_4)_{M+k-\ell}}\nonumber\\
 \hphantom{\frac{h_{\ell, k} }{c^2_{0,0}} =}{} \times \frac{\ell!(1+b_3)_\ell(1+b_4)_M(1+b_1+b_3)}{M!(1+b_1)_\ell(1+b_1+b_3)_\ell(M+3+b_1+b_2+b_3+b_4)_{m}(2\ell+1+b_1+b_3)}.
\end{gather*}

With this normalization, the $\hat{g}_{\ell,k}(s,t)$ form an orthonormal basis.

\subsection{The norm of 1}

Throughout the previous analysis, the weight and normalization have been proportional to an arbitrary overall factor $c_{0,0}$.
 We can f\/ix this constant by requiring that the function~$1$, belonging to the basis $h_{n,k}(s,t)$ of monomials in $s^2$, $t^2$, be
normalized to length~$1$. We compute this by using the Wilson orthogonality for the $0$th order Wilson polynomials (i.e.~(\ref{Racah}) with $k, k'=0$).
The norm of $1$ is given by
\begin{gather*} \langle 1, 1 \rangle  = \sum_{m=0}^M\sum_{\ell=0}^m \omega(t_\ell, s_m)\nonumber \\
\phantom{\langle 1, 1 \rangle}{} =  \sum_{m=0}^M\sum_{\ell=0}^m\frac
{M!(1+b_4)_{M-m}(M+b_1+b_2+b_3+b_4+3)_m(2m+2+b_1+b_2+b_3)}
{(M-m)!(1+b_4)_M(M+b_1+b_2+b_3+3)_m(2+b_1+b_2+b_3)}\nonumber\\
\phantom{\langle 1, 1 \rangle=}{}\times \frac{(1+b_1)_\ell(1+b_1+b_3)_\ell(1+b_2)_{m-\ell}(2+b_1+b_2+b_3)_{m+\ell}(2\ell+1+b_1+b_3)}
{\ell!(m-\ell)!(1+b_3)_\ell(2+b_1+b_3)_{m+\ell}(1+b_1+b_3)c^2_{0,0}} . %\!\!\!\!\label{norm1}
\end{gather*}
Evaluating the double sum gives
\begin{gather*} %\label{norm1 nm}
\langle 1, 1 \rangle =\frac{(3+b_1+b_2+b_4)_M(3+b_1+b_2+b_3)_M}{(1+b_4)_M(1+b_3)_M}\frac1{c^2_{0,0}}.
\end{gather*}

Thus, setting
 \begin{gather*}
 c_{0,0}^2=\frac{(3+b_1+b_2+b_4)_{M}(3+b_1+b_2+b_3)_{M}}{(1+b_4)_M(1+b_3)_{M}},
 \end{gather*}
will make $\langle 1, 1\rangle =1$.

\subsection{Inf\/inite dimensional representations}
For  inf\/inite dimensional but bounded below representations with $-m$ and $-M+\ell$ nonnegative integers and all $b_j$ real
 we take the inner product of two functions $f(t^2,s^2)$, $g(t^2,s^2)$ in the form
\begin{gather*}%\label{2varinnerprod}
\langle f,g\rangle =\int_0^\infty\int_0^\infty f\left(-t^2,-s^2\right)g\left(-t^2,-s^2\right)\omega(t,s)  dt  ds.
\end{gather*}
To compute the measure $\omega(t,s)$ such that our operators $L_{jk}$ are formally self-adjoint we use the fact that we already know the restriction of the measure to the $s$-constant basis (\ref{fbasis}) and the $t$-constant basis (\ref{gbasis}). For consistency,  we see that the weight function should be
\begin{gather*}
\frac{1}{(2\pi)^2}\left|\frac{\Gamma(\frac{b_2+1}{2}+i(t+s))\Gamma(\frac{b_2+1}{2}+i(-t+s))\Gamma(\frac{b_1+b_3+1}{2}+it)
\Gamma(\frac{b_1-b_3+1}{2}+it)}{\Gamma(2it)}\right|^2 \\
\qquad{} \times \left| \frac{\Gamma(-M-\frac{b_1+b_2+b_3}{2}-1+is)\Gamma(M+\frac{b_1+b_2+b_3}{2}+b_4+2+is)}{\Gamma(2is)}\right|^2=\omega(t,s).
\end{gather*}
Then we can compute the norm square of the constant function $f(t^2,s^2)=1$ by using (\ref{Wilsonnorm}) twice to  evaluate the iterated integral:
\begin{gather*}
\langle 1,1\rangle =4\frac{\Gamma(b_1+1)\Gamma(b_2+1)\Gamma(b_4+1)}{\Gamma(b_1+b_2+b_4+3)}\nonumber\\
\phantom{\langle 1,1\rangle =}{} \times  \Gamma(-M)\Gamma(-M-b_3)\Gamma(M+b_1+b_2+b_3+b_4+3)\Gamma(M+b_1+b_2+b_4+3).%\label{2norm}
\end{gather*}

Comparing the measures that we have derived for the inf\/inite dimensional and f\/inite dimensional cases with the
 2-variable Wilson and Racah polynomials introduced by Tratnik~\cite{T1,T2, GI}  we see that they agree.
Thus we have found  two-variable Wilson and Racah polynomials in complete generality.

\section{Expansion coef\/f\/icients}

We can easily determine the  coef\/f\/icients for the expansion of one of our bases in terms of another.
Here we write the expansion coef\/f\/icients in terms of the unnormalized functions.
The expansion of the $d_{\ell,m}$ basis in terms of the $f_{n,m}$ basis is given by
\begin{gather*} \langle \delta(t-t_\ell)\delta(s-s_m),w_{n}(t)\delta(s-s_m')\rangle=\delta_{m,m'} w_{n}(t_\ell)|_{s=s_m}\omega(t_\ell,s_m),\\
  w_{n}(t_\ell)|_{s=s_m}=(-m)_{n} (-m-b_3)_{n} (1+b_2)_{n}\\
\phantom{w_{n}(t_\ell)|_{s=s_m}=}{}
\times {}_4F_3\left( \begin{array}{cccc}-n,& 1+b_1+b_2+n,& -1-b_1-b_3-m-\ell, &-m+\ell\\  -m, & -m-b_3, &  1+b_2,&\end{array} \bigg{|} 1\right).
\end{gather*}

The expansion of the $d_{\ell,m}$ basis in terms of the $g_{\ell,k}$ basis is given by
\begin{gather*} \langle \delta(t-t_\ell)\delta(s-s_m),w_{k}(s)\delta(t-t_\ell)\rangle=\delta_{n,n} w_{m'-n}(s_m)\omega(t_n,s_m),\\
  w_{k}(s_m)|_{t=t_\ell}=(-M+\ell)_{k} (3+b_1+b_2+b_3+b_4+M+\ell)_{k}(1+b_2)_{k} \\
\times  {}_4F_3\left( \!\!\! \begin{array}{cccc}-k,& 1+b_2+b_4+k,& -m+\ell, &2+b_1+b_2+b_3+m+\ell \\  -M+\ell, & 3+b_1+b_2+b_3+b_4+M+\ell,&  1+b_2,&
\end{array} \!\! \bigg{|} 1\right).\!
\end{gather*}

Finally, the expansion of the  $f_{n,m}$ basis in terms of the $g_{\ell,k}$ basis is given by
\begin{gather*}
\langle w_n(t)\delta(s-s_m),w_{k}(s)\delta(t-t_\ell)\rangle= w_{n}(t_n)|_{s=s_m}w_{k}(s_m)|_{t=t\ell}\omega(t_\ell,s_m).
\end{gather*}

In order to understand the signif\/icance of these expansions in quantum theory, it is useful to consider the
results of  \cite{KMT}. There the Schr\"odinger eigenvalue problem for the generic potential on the $n$-sphere was considered,
for general $n$. For $n=3$ it was  shown that all of the eigenfunctions of the pairs commuting operators treated in this paper separated in
some version of either spherical or cylindrical coordinates and were expressible as continuous multivariable orthogonal
polynomials orthogonal on a simplex. Thus the expansion coef\/f\/icients derived here represent the expansion of one basis of solutions of the
Schr\"odinger eigenvalue equation in terms of another.

\subsection[A basis for $L_{12}$, $L_{12}+L_{14}+L_{24}$]{A basis for $\boldsymbol{L_{12}}$, $\boldsymbol{L_{12}+L_{14}+L_{24}}$}

Now that we have computed the measures for our spaces of polynomials from f\/irst principles and established that they agree with those for the Tratnik
generalization of Wilson and Racah polynomials to two variables \cite{T1, T2}, we can make use of known results for the Tratnik case to compute another
 ON basis for our spaces. In an appendix to~\cite{GI} the authors show that the true 2-variable Racah polynomials def\/ined by Tratnik
are simultaneous eigenfunctions of two commuting dif\/ference operators ${\cal L}_1$, ${\cal L}_2$. We will identify these operators with our symmetry algebra and verify
another eigenbasis for our representation space.

We construct the polynomials of Tratnik \cite{T1,T2} and  operators given in \cite{GI} via the def\/initions
\begin{gather*} x_0=0, \qquad x_1=t-\frac{b_1+b_3+1}2, \qquad x_2=-s-1-\frac{b_1+b_2+b_3}2, \qquad x_3=M,\\
  \beta_0=b_3, \qquad \beta_1=b_1+b_3+1,\qquad \beta_2=b_1+b_2+b_3+2, \qquad \beta_3=b_1+b_2+b_3+3.
  \end{gather*}
The original form of the polynomials were given in terms of the Racah polynomials which can be related to the Wilson polynomials via
\begin{gather*}
r_n(a, b, c, d, x)=w_n\big(\tilde{a}, \tilde{b}, \tilde{c}, \tilde{d}, (x+\tilde{a})^2\big)
\end{gather*}
with
\[
\tilde{a} =\frac{c+d+1}2, \qquad \tilde{b}=a-\frac{c+d-1}2,\qquad \tilde{c}=b+\frac{d-c+1}2,\qquad \tilde{d}=\frac{c-d+1}2.
\]
Then,  the two-variable extension of the Wilson polynomials def\/ined by Tratnik are given by equation~(3.10) of~\cite{GI} as
\begin{gather*} R_2 (n_1,n_2;\beta_i; x_1,x_2;M)=r_{n_1}\left(\beta_1-\beta_0-1, \beta_2-\beta_1-1, -x_2-1, x_2+\beta_1; x_1\right)\nonumber\\
\qquad{} \times r_{n_2}\left(n_1+\beta_2-\beta_0-1, \beta_3-\beta_2-1, n_1-M-1, n_1+\beta_2+M; -n_1+x_2\right)
\end{gather*}
with the requirements that $0\le n_1\le n_1+n_2\le M$.

We can express the 2-variable polynomial $R_2$ in terms of the Wilson polynomials using the original parameters and variables of the model as
\begin{gather}\label{R2} R_2 (n_1,n_2;b_i; t,s;M)=w_{n_1}\left(\alpha,\beta,\gamma,\delta; t^2\right)\\
 {} \times w_{n_2}\left(n_1+\beta+\frac{\alpha+\delta}2,\gamma+\frac{\alpha+\delta}2,  M+1+b_4+\beta+\frac{\alpha+\delta}2,M+1+\beta+\frac{\alpha+\delta}2, s^2\right),\nonumber
\end{gather}
where as in (\ref{para1})
\[\alpha=\frac{b_2+1}{2}+s,\qquad \beta =\frac{b_1+b_3+1}{2},\qquad \gamma=\frac{b_1-b_3+1}{2},\qquad \delta=\frac{b_2+1}{2}-s.\]
In particular, note that the parameters $\alpha$, $\delta$ depend on $s$ and so the polynomial $w_{n_1}$ is a function of both~$s$ and~$t$.

Note that it was already demonstrated in Section~\ref{section4.2} that the polynomial, $w_{n_1}$, is an eigenfunction of $L_{12}$. Furthermore, it is easy to see that~$w_{n_2}$ depends only on $s$ and so will be left invariant by~$L_{12}$ and so the 2-variable polynomial $R_2$ is an eigenfuction
for~$L_{12}$. As was exhibited in~\cite{GI}, there is a set of two commuting dif\/ference operators  whose simultaneous eigenfunctions are just
 these orthogonal polynomials. It is then natural to expect that these operators can be expressed in terms of the operators in our
 model which commute with $L_{12}$, i.e.,  ${I}$, $L_{12}$, $L_{34}$,  $H_3$, and  $H_4$.

The  commuting dif\/ference operators are given as follows, via~\cite{GI}. Let $I_i$ be the operator which maps $x_i$ to $-x_i-\beta_i$ and
leaves f\/ixed $x_j$ for $j\ne i$. Similarly, def\/ine  $E_{x_1}^a$ as the operator which maps $x_i$ to $x_i+a$ and leaves f\/ixed  $x_j$ for $j\ne i$.
 We def\/ine functions $B_i^{j,k}$ as
\begin{gather*} B_i^{0,0} \equiv x_i(x_i+\beta_i)+x_{i+1}(x_{i+1}+\beta_{i+1})+\frac{(\beta_1+1)(\beta_{i+1}-1)}2,\\
B_{i}^{0,1} \equiv(x_{i+1}+x_i+\beta_{i+1})(x_{i+1}-x_i+\beta_{i+1}-\beta_i),\\
B_{i}^{1,0} \equiv (x_{i+1}-x_i)(x_{i+1}+x_i+\beta_{i+1}),\\
B_{i}^{1,1} \equiv(x_{i1+}+x_i+\beta_{i+1})(x_{i+1}+x_{i}+\beta_{i+1}+1),
\end{gather*}
and further extend these functions for $k=-1,0,1$ via
\begin{gather*}
B_{i}^{-1,k}\equiv I_i\big(B_i^{1,k}\big), \qquad B_{i}^{k,-1}\equiv I_{i+1}\big(B_i^{k,i}\big).
\end{gather*}
We also def\/ine $b_i^j$ by
\begin{gather*}
b_{i}^0 =(2x_{i}+\beta_{i}+1)(2x_{i}+\beta_i-1),\qquad
b_i^1 =(2x_i+\beta_i+1)(2x_i+\beta_1),\qquad
b_i^{-1} =I_i\big(b_i^1\big). \end{gather*}
Let $\nu$ be some multi-index $\nu=(\nu_1, \nu_2)$ with $\nu_i=-1,0,1$ and $\mu$ be a single index $\mu=-1,0,1$. Then the functions  given by
\begin{gather*} C_\mu\equiv2^{1-|\mu|}\frac{B_0^{0,\mu}B_1^{\mu,0}}{b_1^\mu},\qquad
 C_\nu\equiv 2^{2-|\nu_1|+|\nu_2|}\frac{B_{0}^{0,\nu_1}B_1^{\nu_1,\nu_2}B_2^{\nu_2,0}}{b_1^{\nu_1}b_2^{\nu_2}},
 \end{gather*}
are enough to def\/ine the operators and describe the results of~\cite{GI}. The operators
\begin{gather*} {\cal L}_1 =\sum_{\mu=-1,0,1}C_\mu E_{x_1}^\mu-\left(x_2(x_2+\beta_2)+\frac{(\beta_0+1)(\beta_2-1)}2\right),\\
{\cal L}_2 =\sum_{\nu_i=-1,0,1}C_\nu E_{x_1}^{\nu_1}E_{x_2}^{\nu_2}-\left(x_3(x_3+\beta_3)+\frac{(\beta_0+1)(\beta_3-1)}2\right),
\end{gather*}
commute and their eigenfunctions are given by $R_2$ (\ref{R2}).

The eigenvalues of the operators ${\cal L}_1$, ${\cal L}_2$ are given by
\begin{gather*} {\cal L}_1R_2 =-n_1(n_1+b_1+b_2)R_2,\qquad
{\cal L}_2R_2 =-(n_1+n_2)(n_1+n_2+b_1+b_2+b_4)R_2,
\end{gather*}
 so we hypothesize that ${\cal L}_1$ is a linear combination of $L_{12}$ and the identity and ${\cal L}_2$ is a linear combination of $H_3$
and the identity. In fact, it is straightforward to verify that
\begin{gather*} L_{12}=4{\cal L}_1-2b_1b_2-2b_1-2b_2-\frac32,\\
  L_{14}+L_{12}+L_{24}=4{\cal L}_2-2(b_1b_2-b_2b_4-b_1b_4)-4(b_1+b_2+b_4)-\frac92.
  \end{gather*}
The normalization of this basis can be found in \cite{T2} and \cite{GI}. Thus, we have shown that Tratnik's version of two-variable
Racah polynomials corresponds
to the $L_{12}$, $L_{12}+L_{14}+L_{24}$ eigenbasis.

\section{Conclusions and discussion}
 We have demonstrated explicitly the isomorphism between the quadratic algebra of the generic
quantum superintegrable system on the 3-sphere and the quadratic algebra generated by the recurrence relations for  two-variable
Wilson and Racah polynomials, and have worked out the basic theory for physically relevant (boundstate) inf\/inite as well as f\/inite dimensional
representations of the algebra. The 6 generators of the quadratic algebra break the degeneracy of the energy eigenspaces and,
via various choices of commuting pairs of operators, allow one to describe unique bases. The eigenbases for the commuting pairs $\{L_{12}, H_4\}$,
$\{L_{13}, H_4\}$
and $\{L_{12},H_3\}$ correspond to separation of the original quantum mechanical Schr\"odinger eigenvalue equation in various polyspherical coordinates, whereas the
eigenbasis for $\{L_{13}, L_{24}\}$ corresponds to separation in cylindrical coordinates, see \cite{KMT}.

Natural questions here are: what is the  origin of these models of the symmetry algebra action and how can we determine when there is a dif\/ferential operator model, a dif\/ference operator model or some other model? Clearly, the models are associated with the spectral resolutions of systems of commuting  operators in the symmetry algebra. In~\cite{KMP2008} we showed how dif\/ference and dif\/ferential operator models can be suggested by analysis of the corresponding classical systems, and these  ideas are relevant here. In \cite{KKM2011} we developed a recurrence relation approach for dif\/ferential operators that allowed us to derive dif\/ference equation models for 2D quantum systems and, again, this approach should generalize to 3D quantum systems. Also, there is an obvious connection between the existence of models and bispectrality~\cite{GI}. Another issue is that all models that we know of for quantum symmetry algebras of 2D and 3D superintegrable systems are associated with commuting operators whose simultaneous separated eigenfunctions are of hypergeometric type. Do there exist models with commuting operators whose simultaneous separated eigenfunctions are not hypergeometric?

 It is suggested by our method that  most
of  the quadratic algebras for all St\"ackel
equivalence classes of 3D second order quantum superintegrable systems on conformally f\/lat spaces should  be obtainable by appropriate limit processes from the
quadratic algebra associated with the generic superintegrable system
on the 3-sphere, namely that generated by the two-variable  Wilson
polynomials. However these limit processes are very intricate, see e.g.~\cite{KKWMPOG}, and each
equivalence class exhibits unique structure, so each class is
important for study by itself. Moreover, within each  class of
St\"ackel equivalent systems the structure of the quadratic algebra
remains unchanged but the spectral analysis of the generators for the
algebra can change. We conjecture that this li\-mi\-ting process for superintegrable quantum systems is analogous to the Askey scheme
for obtaining various families of orthogonal polynomials as limits of Askey--Wilson polynomials.

As an example of this,
in the paper~\cite{KMPost10} we studied the quadratic algebra associated with the quantum 3D caged isotropic oscillator.
There, the  Hamiltonian operator was \begin{gather*}%\label{qham3}
H=\partial_1^2+\partial_2^2+\partial_3^2+a^2\left(x_1^2+x_2^2+x_3^2\right)
+\frac{b_1}{x_1^2}+\frac{b_2}{x_2^2}+\frac{b_3}{x_3^2},\qquad
\partial_i\equiv\partial_{x_i},\end{gather*} and a basis for the second order constants of the motion was (with $H =M_1+M_2+M_3$)
\begin{gather*}%\label{q2ordsym3}
M_\ell=\partial_\ell^2+a^2x_\ell^2+\frac{b_\ell}{x_\ell^2},\quad \ell=1,2,3,\qquad
L_i=(x_j\partial_k-x_k\partial_j)^2+\frac{b_jx_k^2}{x_j^2}+\frac{b_kx_j^2}{x_k^2}.\end{gather*} We found 3 two-variable models for
 physically relevant irreducible representations
of the quad\-ra\-tic algebra. One was in terms of dif\/ferential operators and led to monomial eigenfunctions for the generators that
corresponded to
separation of variables in Cartesian coordinates, one was in terms of mixed dif\/ferential-dif\/ference operators and led to one-variable
dual Hahn polynomial eigenfunctions for the generators that corresponded to
separation of variables in cylindrical coordinates, and the third was in terms of pure dif\/ference operators and led to one-variable
Wilson or Racah polynomial eigenfunctions for the generators that corresponded to
separation of variables in spherical coordinates. It can be shown that the f\/lat space caged isotropic oscillator system can be
obtained as a limit of the generic system on the sphere, whereas at the quadratic algebra level one variable dual Hahn and Wilson polynomials can
be obtained as limits of two-variable Wilson polynomials.

For
$n$D  nondegenerate superintegrable systems on conformally f\/lat spaces there are \mbox{$2n-1$} functionally independent
but $n(n+1)/2$ linearly independent generators for the quadratic
algebra. It is reasonable to conjecture that the quadratic algebra of the generic potential on the $n$-sphere is uniquely associated with
the $(n-1)$-variable version of  Tratnik's multivariable Wilson  polynomials.

Finally, these results suggest the existence of  a $q$ version of superintegrability for quantum systems~\cite{GR}.

\appendix

\section{Recurrence relations for Wilson polynomials}\label{AppendixA}

In addition to a three term recurrence relation, the Wilson polynomials $\Phi^{(\alpha,\beta,\gamma,\delta)}_{n}(t^2)$
 satisfy the following parameter-changing recurrence relations:
\begin{gather*}%\label{recurrence1a}
\tau^{(\alpha,\beta,\gamma,\delta)}\Phi^{(\alpha,\beta,\gamma,\delta)}_{n} = \frac{n(n+\alpha+\beta+\gamma+\delta-1)}{(\alpha+\beta)(\alpha+\gamma)(\alpha+\delta)}\Phi^{(\alpha+1/2,\beta+1/2,\gamma+1/2,\delta+1/2)}_{n-1},
\end{gather*}
where
\[
\tau^{(\alpha,\beta,\gamma,\delta)}=\frac{1}{2y}\big(T^{1/2}-T^{-1/2}\big).
\]
This is a consequence of
\begin{gather*}
 \tau[(\alpha+y)_k(\alpha-y)_k]=-k\left[\left(\alpha+\frac12+y\right)_{k-1}\left(\alpha+\frac12-y\right)_{k-1}\right],\nonumber\\
% \label{recurrence2a}
 \mu^{(\alpha,\beta,\gamma,\delta)}\Phi^{(\alpha,\beta,\gamma,\delta)}_{n} = -(\alpha+\beta-1)\Phi^{(\alpha-1/2,\beta-1/2,\gamma+1/2,\delta+1/2)}_{n},
 \end{gather*}
where
\[
 \mu^{(\alpha,\beta,\gamma,\delta)}=\frac{1}{2y}\left[-\left(\alpha+y-\frac12\right)
 \left(\beta+y-\frac12\right)T^{1/2}+\left(\alpha-y-\frac12\right)\left(\beta-y-\frac12\right)T^{-1/2}\right].
 \]
This follows from
\[
 \mu[(\alpha+y)_k(\alpha-y)_k]=-(\alpha+\beta-k-1)\left[\left(\alpha-\frac12+y\right)_{k}\left(\alpha-\frac12-y\right)_{k}\right].
 \]

Using the inner product (\ref{Wilsonnorm}), one can compute the adjoint recurrences
\begin{gather*}%\label{recurrence3a}
\tau^{*(\alpha+1/2,\beta+1/2,\gamma+1/2,\delta+1/2)}\Phi^{(\alpha+1/2,\beta+1/2,\gamma+1/2,\delta+1/2)}_{n-1} = (\alpha+\beta)(\alpha+\gamma)(\alpha+\delta)\Phi^{(\alpha,\beta,\gamma,\delta)}_{n},
\end{gather*}
where
\begin{gather*}
 \tau^{*(\alpha+1/2,\beta+1/2,\gamma+1/2,\delta+1/2)}\\
 \qquad{}=
 \frac{1}{2y}\big[(\alpha+y)(\beta+y)(\gamma+y)(\delta+y)T^{1/2}-(\alpha-y)(\beta-y)(\gamma-y)(\delta-y)T^{-1/2}\big],
 \end{gather*}
and
\begin{gather*}%\label{recurrence4a}
\mu^{(\gamma+1/2,\delta+1/2,\alpha-1/2,\beta-1/2)}\Phi^{(\alpha-1/2,\beta-1/2,\gamma+1/2,\delta+1/2)}_{n}  \\
\qquad{}
 = -\frac{(n+\gamma+\delta)(n+\alpha+\beta-1)}{(\alpha+\beta-1)}\Phi^{(\alpha,\beta,\gamma,\delta)}_{n},
 \end{gather*}
which follows from
\begin{gather*}
 \mu^{(\gamma+1/2,\delta+1/2,\alpha-1/2,\beta-1/2)}\left[\left(\alpha-\frac12 +y\right)_k\left(\alpha-\frac12-y\right)_k\right]\\
 \qquad{}=
 -(k+\gamma+\delta)[ (\alpha+y)_k(\alpha-y)_k]\\
 \qquad\quad {} +k    (\alpha+\gamma+k-1)(\alpha+\delta+k-1)[(\alpha+y)_{k-1}(\alpha-y)_{k-1}].
 \end{gather*}

Due to the symmetry in $\alpha$, $\beta$, $\gamma$, $\delta$, the $\mu$ operators lead to several recurrences. Basically they allow one to raise any two of the parameters by $1/2$ and lower the remaining two parameters by~$1/2$ in such a way that the process is essentially reversible. Moreover, composing a recurrence with its adjoint leads to eigenvalue equations for the Wilson polynomials.

\section{Recurrence relations for construction of the  spherical\\ and cylindrical models}\label{AppendixB}

The spherical and cylindrical models are  associated with the 8 basic raising and lowering ope\-ra\-tors for the Wilson polynomials, as well as the three term recurrence relation.
 We list these operators here and describe their actions on  the basis polynomials $\Phi_n\equiv \Phi^{(\alpha,\beta,\gamma,\delta)}_{n}$.
\begin{alignat*}{3}
& 1. \ \ && R=\frac{1}{2y}\big[T^{1/2}-T^{-1/2}\big],& \nonumber\\
& && R\Phi_{n} =  \frac{n(n+\alpha+\beta+\gamma+\delta-1)}{(\alpha+\beta)(\alpha+\gamma)(\alpha+\delta)}\Phi^{(\alpha+1/2,\beta+1/2,\gamma+1/2,\delta+1/2)}_{n-1}.&
%\label{RW}
\\
 & 2. \ \ && L=\frac{1}{2y}\left[(\alpha-1/2+y)(\beta-1/2+y)(\gamma-1/2+y)(\delta-1/2+y)T^{1/2}\right.& \nonumber\\
 &&& \left. \phantom{L=}{} -(\alpha-1/2-y)(\beta-1/2-y)(\gamma-1/2-y)(\delta-1/2-y)T^{-1/2}\right],& \nonumber\\
&&&  L\Phi_{n} =  (\alpha+\beta-1)(\alpha+\gamma-1)(\alpha+\delta-1)\Phi^{(\alpha-1/2,\beta-1/2,\gamma-1/2,\delta-1/2)}_{n+1}.&
% \label{LW}
\\
& 3. \ \ && L_{\alpha\beta}=\frac{1}{2y}\left[-(\alpha-1/2+y)(\beta-1/2+y)T^{1/2}+(\alpha-1/2-y)(\beta-1/2-y)T^{-1/2}\right], & \nonumber\\
&&&  L_{\alpha\beta}\Phi_{n}  =  -(\alpha+\beta-1)\Phi^{(\alpha-1/2,\beta-1/2,\gamma+1/2,\delta+1/2)}_{n}.&
%\label{Lalphabeta}
\\
& 4. \ \  && R^{\alpha\beta}=\frac{1}{2y}\left[-(\gamma-1/2+y)(\delta-1/2+y)T^{1/2}+(\gamma-1/2-y)(\delta-1/2-y)T^{-1/2}\right], & \nonumber\\
&&&   R^{\alpha\beta}\Phi_{n}  =  -\frac{(n+\gamma+\delta-1)(n+\alpha+\beta)}{\alpha+\beta}\Phi^{(\alpha+1/2,\beta+1/2,\gamma-1/2,\delta-1/2)}_{n}. &
 %\label{Ralphabeta}
 \\
& 5. \ \  && L_{\alpha\gamma}=\frac{1}{2y}\left[-(\alpha-1/2+y)(\gamma-1/2+y)T^{1/2}+(\alpha-1/2-y)(\gamma-1/2-y)T^{-1/2}\right], & \nonumber\\
&&& L_{\alpha\gamma}\Phi_{n}  =  -(\alpha+\gamma-1)\Phi^{(\alpha-1/2,\beta+1/2,\gamma-1/2,\delta+1/2)}_{n}.&
 %\label{Lalphagamma}
 \\
& 6. \ \  && R^{\alpha\gamma}=\frac{1}{2y}\left[-(\beta-1/2+y)(\delta-1/2+y)T^{1/2}+(\beta-1/2-y)(\delta-1/2-y)T^{-1/2}\right], & \nonumber\\
&&&  R^{\alpha\gamma}\Phi_{n}  =  -\frac{(n+\beta+\delta-1)(n+\alpha+\gamma)}{\alpha+\gamma}\Phi^{(\alpha+1/2,\beta-1/2,\gamma+1/2,\delta-1/2)}_{n}.&
%\label{Ralphagamma}
\\
& 7. \ \  &&  L_{\alpha\delta}=\frac{1}{2y}\left[-(\alpha-1/2+y)(\delta-1/2+y)T^{1/2}+(\alpha-1/2-y)(\delta-1/2-y)T^{-1/2}\right], & \nonumber\\
&&& L_{\alpha\delta}\Phi_{n}  =  -(\alpha+\delta-1)\Phi^{(\alpha-1/2,\beta+1/2,\gamma+1/2,\delta-1/2)}_{n}.&
 %\label{Lalphadelta}
 \\
& 8. \ \ && R^{\alpha\delta}=\frac{1}{2y}\left[-(\beta-1/2+y)(\gamma-1/2+y)T^{1/2}+(\beta-1/2-y)(\gamma-1/2-y)T^{-1/2}\right], & \nonumber\\
&&&  R^{\alpha\delta}\Phi_{n}  =  -\frac{(n+\beta+\gamma-1)(n+\alpha+\delta)}{\alpha+\delta}\Phi^{(\alpha+1/2,\beta-1/2,\gamma-1/2,\delta+1/2)}_{n}.&
 %\label{Ralphadelta}
\end{alignat*}
Again, the three term recurrence is
\begin{gather}\label{Wilsonrecurrence1}
y^2\Phi_n\left(y^2\right)=K(n+1,n)\Phi_{n+1}\left(y^2\right)+K(n,n)
\Phi_n\left(y^2\right)+K(n-1,n)\Phi_{n-1}\left(y^2\right),
\end{gather}
where the coef\/f\/icients are given by (\ref{KWilson1}), (\ref{KWilson2}) and  (\ref{KWilson3}).

To construct the spherical model we look for  f\/irst order dif\/ference operators def\/ining recurrences for the basis functions $\Phi_n$   that are not just squares and that change parameters by integer amounts. Here are the possibilities:

1.  Eigenvalue equations.  Here $\alpha$, $\beta$, $\gamma$, $\delta$, $n$ must be unchanged. The possibilities are
\begin{gather*}
 LR\Phi_n=n(n+\alpha+\beta+\gamma+\delta-1)\Phi_n,\\
 L_{\alpha\beta}R^{\alpha\beta}\Phi_n=(n+\gamma+\delta-1)(n+\alpha+\beta)\Phi_n,\\
 L_{\alpha\gamma}R^{\alpha\gamma}\Phi_n=(n+\beta+\delta-1)(n+\alpha+\gamma)\Phi_n,\\ L_{\alpha\delta}R^{\alpha\delta}\Phi_n=(n+\beta+\gamma-1)(n+\alpha+\delta)\Phi_n.
 \end{gather*}

2. Fix $n$ and lower $\alpha$ by 1.  The only possibilities are
\begin{gather*}
 L_{\alpha\beta}L_{\alpha\gamma} \Phi_n^{(\alpha,\beta,\gamma,\delta)}=(\alpha+\beta-1)(\alpha+\gamma-1)\Phi_n^{(\alpha-1,\beta,\gamma,\delta+1)},\\
  L_{\alpha\beta}L_{\alpha\delta} \Phi_n^{(\alpha,\beta,\gamma,\delta)}=(\alpha+\beta-1)(\alpha+\delta-1)\Phi_n^{(\alpha-1,\beta,\gamma+1,\delta)},\\
 L_{\alpha\gamma}L_{\alpha\delta} \Phi_n^{(\alpha,\beta,\gamma,\delta)}=(\alpha+\gamma-1)(\alpha+\delta-1)\Phi_n^{(\alpha-1,\beta+1,\gamma,\delta)}.
\end{gather*}

3. Fix $n$ and raise by $\alpha$ by 1. The only possibilities are
\begin{gather*}
 R^{\alpha\beta}R^{\alpha\gamma} \Phi_n^{(\alpha,\beta,\gamma,\delta)}
 =\frac{(n+\alpha+\beta)(n+\alpha+\gamma)(n+\beta+\delta-1)(n+\gamma+\delta-1)}{(\alpha+\beta)(\alpha+\gamma)}\Phi_n^{(\alpha+1,\beta,\gamma,\delta-1)},\\
  R^{\alpha\beta}R^{\alpha\delta} \Phi_n^{(\alpha,\beta,\gamma,\delta)}
  =\frac{(n+\alpha+\beta)(n+\alpha+\delta)(n+\beta+\gamma-1)(n+\gamma+\delta-1)}{(\alpha+\beta)(\alpha+\delta)}\Phi_n^{(\alpha+1,\beta,\gamma-1,\delta)},\\
  R^{\alpha\gamma}R^{\alpha\delta} \Phi_n^{(\alpha,\beta,\gamma,\delta)}
  =\frac{(n+\alpha+\gamma)(n+\alpha+\delta)(n+\beta+\gamma-1)(n+\beta+\delta-1)}{(\alpha+\gamma)(\alpha+\delta)}\Phi_n^{(\alpha+1,\beta-1,\gamma,\delta)}.
  \end{gather*}

4. Fix $n$ and  $\alpha$. The only possibilities that change parameters are
\begin{gather*}
L_{\alpha\beta}R^{\alpha\gamma}\Phi_n^{(\alpha,\beta,\gamma,\delta)}
=\frac{(\alpha+\beta-1)(n+\alpha+\gamma)(n+\beta+\delta-1)}{\alpha+\gamma}\Phi_n^{(\alpha,\beta-1,\gamma+1,\delta)},\\
 L_{\alpha\beta}R^{\alpha\delta}\Phi_n^{(\alpha,\beta,\gamma,\delta)}
=\frac{(\alpha+\beta-1)(n+\alpha+\delta)(n+\beta+\gamma-1)}{\alpha+\delta}\Phi_n^{(\alpha,\beta-1,\gamma,\delta+1)},\\
 L_{\alpha\gamma}R^{\alpha\beta}\Phi_n^{(\alpha,\beta,\gamma,\delta)}
=\frac{(\alpha+\gamma-1)(n+\alpha+\beta)(n+\gamma+\delta-1)}{\alpha+\beta}\Phi_n^{(\alpha,\beta+1,\gamma-1,\delta)},\\
 L_{\alpha\gamma}R^{\alpha\delta}\Phi_n^{(\alpha,\beta,\gamma,\delta)}
 =\frac{(\alpha+\gamma-1)(n+\alpha+\delta)(n+\gamma+\beta-1)}{\alpha+\delta}\Phi_n^{(\alpha,\beta,\gamma-1,\delta+1)},\\
 L_{\alpha\delta}R^{\alpha\beta}\Phi_n^{(\alpha,\beta,\gamma,\delta)}
=\frac{(\alpha+\delta-1)(n+\alpha+\beta)(n+\gamma+\delta-1)}{\alpha+\beta}\Phi_n^{(\alpha,\beta+1,\gamma,\delta-1)},\\
 L_{\alpha\delta}R^{\alpha\gamma}\Phi_n^{(\alpha,\beta,\gamma,\delta)}
 =\frac{(\alpha+\delta-1)(n+\alpha+\gamma)(n+\beta+\delta-1)}{\alpha+\gamma}\Phi_n^{(\alpha,\beta,\gamma+1,\delta-1)}.
 \end{gather*}

5. Lower $n$  by $1$ and  f\/ix $\alpha$. The only possibilities that change parameters are
\begin{gather*}
RL_{\alpha\beta}\Phi_n =-\frac{n(n+\alpha+\beta+\gamma+\delta-1)}{(\alpha+\gamma)(\alpha+\delta)}\Phi_{n-1}^{(\alpha,\beta,\gamma+1,\delta+1)},\\
RL_{\alpha\gamma}\Phi_n =-\frac{n(n+\alpha+\beta+\gamma+\delta-1)}{(\alpha+\beta)(\alpha+\delta)}\Phi_{n-1}^{(\alpha,\beta+1,\gamma,\delta+1)},\\
RL_{\alpha\delta}\Phi_n =-\frac{n(n+\alpha+\beta+\gamma+\delta-1)}{(\alpha+\beta)(\alpha+\gamma)}\Phi_{n-1}^{(\alpha,\beta+1,\gamma+1,\delta)}.
\end{gather*}

6. Lower $n$  by $1$ and  raise $\alpha$ by $1$. The only possibilities  are
\begin{gather*}
RR^{\alpha\beta}\Phi_n=-\frac{n(n+\gamma+\delta-1)(n+\alpha +\beta)(n+\alpha+\beta+\gamma+\delta-1)}{(\alpha+\beta)(\alpha+\beta+1)(\alpha+\gamma)(\alpha+\delta)}\Phi_{n-1}^{(\alpha+1,\beta+1,\gamma,\delta)},\\
 RR^{\alpha\gamma}\Phi_n=-\frac{n(n+\beta+\delta-1)(n+\alpha +\gamma)(n+\alpha+\beta+\gamma+\delta-1)}{(\alpha+\gamma)(\alpha+\gamma+1)(\alpha+\beta)(\alpha+\delta)}\Phi_{n-1}^{(\alpha+1,\beta,\gamma+1,\delta)},\\
  RR^{\alpha\delta}\Phi_n=-\frac{n(n+\beta+\gamma-1)(n+\alpha +\delta)(n+\alpha+\beta+\gamma+\delta-1)}{(\alpha+\delta)(\alpha+\delta+1)(\alpha+\beta)(\alpha+\gamma)}\Phi_{n-1}^{(\alpha+1,\beta,\gamma,\delta+1)}.
 \end{gather*}

7. Raise $n$  by $1$ and  f\/ix $\alpha$. The only possibilities that change parameters are
\begin{gather*}
R^{\alpha\beta}L\Phi_n=-(n+\gamma+\delta-2)(n+\alpha+\beta-1)(\alpha+\gamma-1)(\alpha+\delta-1)\Phi_{n+1}^{(\alpha,\beta,\gamma-1,\delta-1)},\\
 R^{\alpha\gamma}L\Phi_n=-(n+\beta+\delta-2)(n+\alpha+\gamma-1)(\alpha+\beta-1)(\alpha+\delta-1)\Phi_{n+1}^{(\alpha,\beta-1,\gamma,\delta-1)},\\
 R^{\alpha\delta}L\Phi_n=-(n+\beta+\gamma-2)(n+\alpha+\delta-1)(\alpha+\beta-1)(\alpha+\gamma-1)\Phi_{n+1}^{(\alpha,\beta-1,\gamma-1,\delta)}.
 \end{gather*}

8. Raise $n$  by $1$ and  lower $\alpha$ by $1$. The only possibilities  are
\begin{gather*}
L_{\alpha\beta}L\Phi_n=-(\alpha+\beta-1)(\alpha+\beta-2)(\alpha+\gamma-1)(\alpha+\delta-1)\Phi_{n+1}^{(\alpha-1,\beta-1,\gamma,\delta)},\\
L_{\alpha\gamma}L\Phi_n=-(\alpha+\gamma-1)(\alpha+\gamma-2)(\alpha+\beta-1)(\alpha+\delta-1)\Phi_{n+1}^{(\alpha-1,\beta,\gamma-1,\delta)},\\
L_{\alpha\delta}L\Phi_n=-(\alpha+\delta-1)(\alpha+\delta-2)(\alpha+\beta-1)(\alpha+\gamma-1)\Phi_{n+1}^{(\alpha-1,\beta,\gamma,\delta-1)}.
\end{gather*}
In addition we can use the three term recurrence (\ref{Wilsonrecurrence1}) and multiplication by the operator $y^2$ to both raise and lower $n$ by 1 while f\/ixing the other parameters.

\subsection*{Acknowledgements}
Thanks to Jonathan Kress for valuable advice on computer verif\/ication of the dif\/ference operator realization for
the structure formulas.  S.P.~acknowledges a postdoctoral IMS fellowship awarded by the Mathematical Physics Laboratory of the Centre de Recherches Math\'ematiques.

\pdfbookmark[1]{References}{ref}
\LastPageEnding

\end{document}